\documentclass{article}

\usepackage[a4paper, total={6in, 8in}]{geometry}
\usepackage[utf8]{inputenc}
\usepackage[T1]{fontenc}
\usepackage{hyperref}       
\usepackage{url}            
\usepackage{booktabs}       
\usepackage{amsfonts}       
\usepackage{nicefrac}       
\usepackage{microtype}      
\usepackage{lipsum}
\usepackage{fancyhdr}       
\usepackage{graphicx}       
\usepackage{natbib}
\usepackage{amsmath}
\graphicspath{{media/}}     
\defcitealias{bretherton2017understanding}{BB17}
\defcitealias{vogel2020influence}{Vogel et al, 2020}
\defcitealias{narenpitak2021sugar}{Narenpitak et al. 2021}

\title{Non-precipitating shallow cumulus convection is intrinsically unstable to length-scale growth
\thanks{This preprint is intended for publication in a scientific journal, but has not been peer-reviewed. Copyright and all rights therein are maintained by the Authors or by other copyright owners. It is understood that all persons copying this information will adhere to the terms and constraints invoked by each Author's copyright.}
}

\author{Martin Janssens\textsuperscript{12*}, 
Jordi Vil\`a-Guerau de Arellano\textsuperscript{1}, \\
Chiel C. van Heerwaarden\textsuperscript{1}, 
Stephan R. de Roode\textsuperscript{2}, \\
A. Pier Siebesma\textsuperscript{23}, 
Franziska Glassmeier\textsuperscript{2}}

\date{\small \textsuperscript{1} Meteorology \& Air Quality Department, Wageningen University \& Research \\
\textsuperscript{2} Geoscience \& Remote Sensing Department, Delft University of Technology \\
\textsuperscript{3} Royal Netherlands Meteorological Institute \\
\today}

\begin{document}
\maketitle

\begin{abstract}
    \noindent Condensation in cumulus clouds plays a key role in structuring the mean, non-precipitating trade-wind boundary layer. Here, we summarise how this role also explains the spontaneous growth of mesoscale ($>O(10)$ km) fluctuations in clouds and moisture around the mean state in a minimal-physics, large-eddy simulation of the undisturbed period during BOMEX on a large ($O(100)$ km) domain. Small, spatial anomalies in latent heating in cumulus clouds, which form on top of small moisture fluctuations, give rise to circulations that transport moisture, but not heat, from dry to moist regions, and thus reinforce the latent heating anomaly. We frame this positive feedback as a linear instability in mesoscale moisture fluctuations, whose time-scale depends only on i) a vertical velocity scale and ii) the mean environment's vertical structure. In our minimal-physics setting, we show both ingredients are provided by the shallow cumulus convection itself: It is intrinsically unstable to length scale growth. The upshot is that energy released by clouds at kilometre scales may play a more profound and direct role in shaping the mesoscale trade-wind environment than is generally appreciated, motivating further research into the mechanism's relevance.
\end{abstract}

\section{Introduction}

Shallow clouds organised into mesoscale patterns by convective instabilities have been recognised as a ubiquitous feature of the subtropical marine boundary layer since satellite imagery in the 1960s first revealed them \citep{agee1973review}. While their discovery sparked much research on the role of convective instabilities in patterning boundary layer clouds, much of that research was long focused towards open and closed convective cells \citep[e.g.][]{fiedler1985mesoscale,muller1996three}. Yet, a rich spectrum of cloud patterns can be found outside the paradigm of such mesoscale cellular convection \citep{wood2006spatial}, including for shallow cumulus clouds that top the trade-wind marine boundary layer \citep{stevens2020sugar,denby2020discovering,janssens2021cloud}.


The interest in the self-organisation of trade-wind cumulus has risen in recent years, in response to cloud-resolving simulations of deep convection \citep{muller2012detailed}, which spontaneously develop mesoscale fluctuations in their cloud structures. Since deep convective organisation plays an important role in regulating radiative heat loss from the atmosphere \citep{tobin2012observational}, it seemed natural to ask whether the observed shallow convective organisation plays a similarly important role. \citet{bony2020sugar} suggest that the answer to this question is yes; in observations, different trade-wind cumulus patterns, forming under different larger-scale conditions, have different cloud feedbacks. Given the disparity between observations and climate model simulations of the trade-cumulus feedback \citep{myers2021observational,cesana2021observational}, this provides ample motivation for better understanding the processes that pattern shallow cumulus-topped marine boundary layers.

Many mesoscale cumulus patterns may simply be either passive responses to mesoscale heterogeneities in cloud-controlling conditions driven by larger-scale dynamics, or are remnants of extratropical disturbances advected into the trades \citep{schulz2021characterization}. However, several others appear to result from the shallow convection itself. Sub-cloud layer rain evaporation can trigger density currents that force new convection upon collision \citep{seifert2013large,zuidema2017survey}, while heterogeneous radiative cooling can drive circulations that lead to cloud clustering \citep{klinger2017effects,naumann2019moist}. Even simulations of clear convective boundary layers \citep{jonker1999mesoscale} and stratocumulus-topped layers \citep{de2004large} spontaneously develop appreciable mesoscale fluctuations in their moisture fields. Building on these studies, \citet{bretherton2017understanding} (\citetalias{bretherton2017understanding} hereafter) in a remarkably thorough piece of work noted that even non-precipitating shallow cumulus convection - stripped of all interactive precipitation and radiation feedbacks - self-organises into clusters in large-eddy simulations (LESs) on domains larger than 100 km, over similar time scales as observed in nature \citep{narenpitak2021sugar}.






This finding is rather striking, since studies of the slab-averaged structure of such layers have a rich history, dating back to the observational budget surveys of the Northeast Pacific \citep{riehl1951north}, the Atlantic Tradewind Experiment \citep[ATEX,][]{augstein1973mass} and the Barbados Oceanographic and Meteorological Experiment \citep[BOMEX,][]{nitta1974heat}. These campaigns established the classical picture of the trade-wind boundary layer: It is maintained by transport of liquid water into the trade-inversion, where its evaporation moistens and cools the layer sufficiently to balance the drying and heating from the subsiding environment. Using data from the BOMEX campaign, \citet{betts1973non,betts1975parametric} then traced the energy supporting the fluxes of liquid water into the inversion to the slab-averaged net condensation in the conditionally unstable cloud layer below. 



Here, we will revisit minimal-physics LES of the ``undisturbed'' period during phase 3 of BOMEX, detailed in section \ref{s:num}, with two objectives. First, we will review \citetalias{bretherton2017understanding}'s instability in this classical view of the the trade-wind layer, to show that the instability can be understood as a natural extension of the role played by net condensation in the slab mean (section~\ref{s:ana}) to mesoscale fluctuations around that mean (section~\ref{s:mec}). Our hope is that making this connection will aid the interpretative side of future studies of the mechanism, e.g. in attempts to understand the relative importance of processes that can pattern trade-wind clouds.

Our second objective is to study the origins of the instability more quantitatively than \citetalias{bretherton2017understanding}. To do so, we extend their theory to a linear stability model for bulk mesoscale moisture fluctuations, and examine its conditions for instability (section~\ref{s:add}). We will show that these are satisfied by the cumulus convection itself, and do not require anything from the large-scale environment, other than that it supports a cumulus layer. Put differently, we will conclude that shallow cumulus convection is intrinsically unstable to length scale growth. 
We end the paper by discussing the relevance of these findings to several ongoing studies of the self-organising cumulus layer, and suggest a few directions that such future research could take (section~\ref{s:dis}). A summmary is given in section~\ref{s:sum}.



\section{Large-eddy simulation of the undisturbed period during BOMEX}
\label{s:num}

\subsection{Case study}

We consider a situation based on observations performed on 22 and 23 June 1969, during phase 3 of BOMEX. There are many reasons for this. First, during this so-called ``undisturbed'' period, the vertical slab-mean moisture and heat profiles were observed to be in a nearly steady state, capped by a well-defined inversion. In fact, the steadiness of these days was an important reason to select them for the budget studies that diagnosed the main features of cumulus convection in an undisturbed environment \citep{holland1973measurements,nitta1974heat}. Later, this also helped popularise the case as a testbed for validating LES models \citep{siebesma1995evaluation,siebesma2003large}, as it allowed comparing statistics averaged over long time-periods. As a result, the undisturbed period is perhaps the single most most studied realisation of the trade-wind boundary layer. All these features make the situation attractive for our study, since it is our objective to use LES to study the development of fluctuations around a steady mean state, departing from the well-established theory from the early observational work. 

It is worth pausing here to note that \citet{nitta1974heat} already show that the trade-wind layer is usually not in steady state, but is highly variable. Furthermore, recent observations of the subtropical Atlantic reveal that the trades usually feature stronger winds, weaker subsidence and stronger temperature inversions than observed during the undisturbed period, often associated with larger-scale, precipitating cloud structures \citep{schulz2021characterization}. Therefore, the situation we study should be considered illustrative, rather than representative.

The second reason we concentrate on BOMEX is that \citetalias{bretherton2017understanding} also report LES results of the case. Their simulations produce significant mesoscale moisture and cloud fluctuations, if run for several days on domains whose horizontal dimensions exceed 100 km $\times$ 100 km. Hence, we will be able to translate rather directly between their results and ours.

Finally, the BOMEX setup we consider excludes and simplifies a number of processes. Of particular interest here are that the case i) ignores spatial and temporal variability in the large-scale subsidence, horizontal wind and surface fluxes of heat and moisture, instead imposing constant forcings for all three, ii) does not locally calculate radiative heating rates, approximating them with a slab-averaged cooling, and iii) explicitly ignores the formation and impact of precipitation. This will suppress length scale growth encouraged by large-scale vertical ascent \citep{narenpitak2021sugar}, radiation \citep{klinger2017effects} and cold pools \citep{seifert2013large}, respectively, all of which appear to be important pathways to develop the mesoscale cumulus patterns observed in nature. 

We do not suggest that variable larger-scale forcing, radiation and precipitation do not influence the length scale growth in shallow cumulus fields. We merely note that \citetalias{bretherton2017understanding} find that they are not necessary ingredients; they merely act to modulate an internal, dynmical growth mechanism that also occurs without them. The mechanism in question is thus fundamentally rooted in moist, shallow convection, and its understanding is clarified by only studying this aspect. 

\subsection{Model setup}

We simulate BOMEX using the Dutch Atmospheric Large Eddy Simulaton \citep[DALES,][]{heus2010formulation,ouwersloot2017large}. We run the case precisely as reported by \citet{siebesma2003large}, save for its computational grid, integration time and advection scheme. To allow the formation of mesoscale fluctuations with little influence from the finite domain size, the cases are run on horizontally square domains spanning 102.4 km, with a height of 10 km, for 36 hours. The horizontal grid spacing $\Delta x = \Delta y = 200$ m, while the vertical grid spacing $\Delta z = 40$ m up to 6 km; it is stretched by 1.7\% per level above this height. The case is run with a variance-preserving, second-order central difference scheme to represent advective transfer. We will concentrate our analysis on the early phase of the simulation, since it develops strong moisture fluctuations that approach the scale of the domain length after around 18 hours. Subsequently, deep convective clouds develop. Such situations are deemed unrealistic in our non-precipitating simulations over domains with doubly-periodic boundary conditions.

\section{The classical theory}
\label{s:ana}

In the anelastic approximation adopted by our LES code, the local budget of a generic scalar $\chi$, which here will denote water or a measure of heat, can be written as

\begin{equation}
    \frac{\partial \chi}{\partial t} = -\frac{\partial}{\partial x_{jh}}\left(u_{jh}\chi\right)
    -\frac{1}{\rho_0}\frac{\partial}{\partial z}\left(\rho_0 w\chi\right)+ S_{\chi},
    \label{e:localbudg}
\end{equation}

\noindent where $u_{jh}$ contains the horizontal velocity vector, the subscript $jh$ indicates summation over the horizontal coordinate, $w$ is the vertical velocity, $\rho_0(z)$ is a profile of reference density and $S_\chi$ is a local source. Since we are interested in fluctuations in $\chi$, we introduce the definition

\begin{equation}
    \chi = \overline{\chi} + \chi',
    \label{e:fluc}
\end{equation}

\noindent where $\overline{\chi}$ and $\chi'$ respectively refer to the slab-average and resulting fluctuation of any model variable. It is instructive to use this partitioning to rewrite eq.~\ref{e:localbudg} on the following form (see appendix A for a step-by-step procedure):

\begin{multline}
    \frac{\partial \chi}{\partial t} = -\overline{u_{jh}}\frac{\partial \chi}{\partial x_{jh}}
    -u_{jh}'\frac{\partial \overline{\chi}}{\partial x_{jh}}
    -\frac{\partial}{\partial x_{jh}}\left(u_{jh}'\chi'\right)
    -\overline{w}\frac{\partial \chi}{\partial z}
    -w'\frac{\partial \overline{\chi}}{\partial z}
    -\frac{1}{\rho_0}\frac{\partial}{\partial z}\left(\rho_0 w'\chi'\right)+ S_{\chi},
    \label{e:localbudgrew}
\end{multline}

\noindent where we have used the anelastic conservation of mass. Eq.~\ref{e:localbudgrew} will serve as our point of departure for the rest of the study. In it, the first three terms describe horizontal transport i) with the mean wind, ii) with fluctuations in the horizontal velocity, and iii) through turbulent fluxes, respectively. The fourth term represents transport with the mean vertical velocity, often associated with the prevailing subsidence of the trades, the fifth term denotes transport with vertical velocity fluctuations against the mean gradient, the sixth term describes vertical turbulent transport, and the final term is again reserved for sources.

\subsection{Slab-averaged heat and moisture budgets}

We will first briefly summarise the dynamics that govern the slab-averaged thermodynamic structure, since these dynamics turn out to also mostly explain the development of mesoscale fluctuations on top of it. The mean scalar budgets can be derived from eq.~\ref{e:localbudgrew} by i) slab-averaging it over a sufficiently large region to represent an ensemble average and ii) assuming the horizontal flux divergence out of the region over which we average is small: 

\begin{multline}
    \underbrace{
    \frac{\partial \overline{\chi}}{\partial t}}_{\text{Tendency}} + \underbrace{\overline{u_{jh}}\frac{\partial \overline{\chi}}{\partial x_{jh}}}_{\text{Horizontal advection}}
    +\underbrace{\overline{w}\frac{\partial \overline{\chi}}{\partial z}}_{\text{Subsidence}} =
    -\underbrace{\frac{1}{\rho_0}\frac{\partial}{\partial z}\left(\rho_0 \overline{ w'\chi'}\right)}_{\text{Vertical flux convergence}}+ \underbrace{\overline{S_{\chi}}}_{\text{Sources}} \equiv \underbrace{Q}_{\text{Apparent source}},
    \label{e:slabav}
\end{multline}

$Q$ defines \citet{yanai1973determination}'s apparent heat source and moisture sink, if the equations are posed for appropriate heat and moisture variables, respectively. We will present these budgets for the two prognostic variables in our LES model, which are conserved under non-precipitating cumulus convection: The total water specific humidity $q_t$ and liquid-water potential temperature $\theta_l$, approximated as

\begin{equation}
    \theta_l \approx \theta - \frac{L_v}{c_p\Pi}q_l.
    \label{e:thl}
\end{equation}

Here, $\theta=T/\Pi$ is (dry) potential temperature, $T$ is temperature, $L_v$ is the latent heat of vaporisation, $c_p$ is the specific heat of dry air at constant pressure, $q_l$ is the liquid water specific humidity and

\begin{equation}
    \Pi = \left(\frac{p}{p_0}\right)^{\frac{R_d}{c_p}}
\end{equation}

\noindent is the Exner function, with $R_d$ the gas constant of dry air, $p$ the reference pressure profile, and $p_0 = 10^{5}$ Pa. 

\begin{figure}
    \centering
    \includegraphics[width=\textwidth]{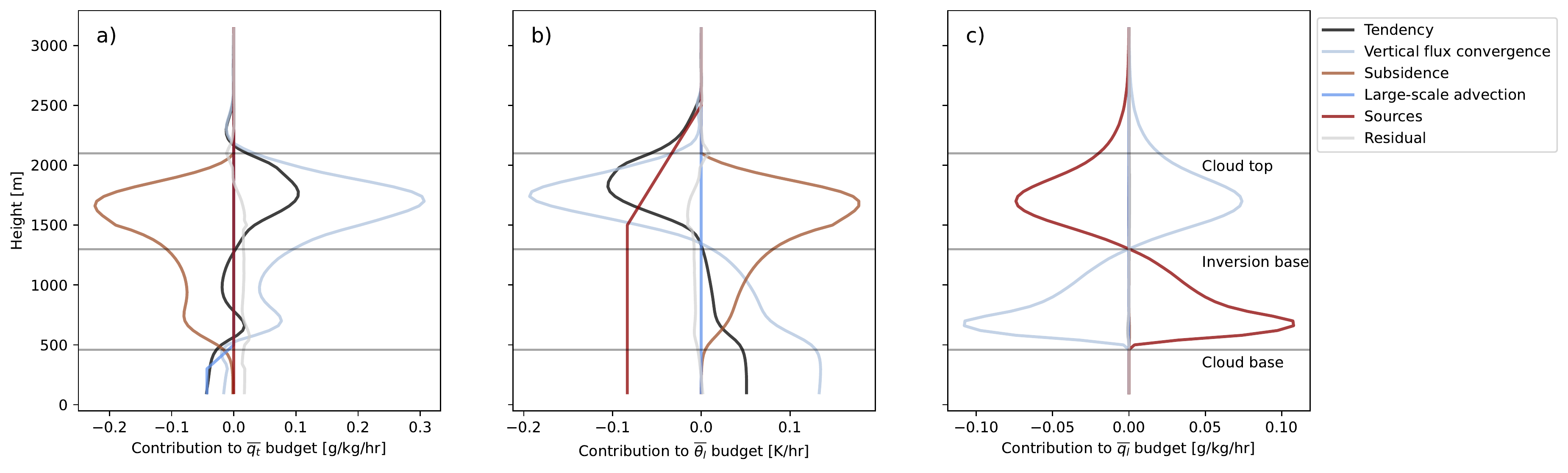}
    \caption{Slab-averaged contributions to the slab-averaged tendencies of $q_t$ (a), $\theta_l$ (b) and $q_l$ (c) (eq.~\ref{e:slabav} rewritten for $\partial\overline{\chi}/\partial t$ on the left-hand side), additionally averaged over hours 2-16 of the simulation. The source of $\overline{\theta_l}$ is radiative, negative and imposed, the source of $\overline{q_l}$ is the net condensation rate.}
    \label{f:slabav}
\end{figure}

In our LES model, which features doubly periodic boundary conditions, slab averages taken of horizontal gradients and vertical velocity are zero by definition. Therefore, we impose the horizontal transport and the vertical velocity in the subsidence term on the left-hand side of eq.~\ref{e:slabav}, in addition to a slab-averaged radiative cooling sink in the budget for $\theta_l$. The resulting contributions to eq.~\ref{e:slabav} are plotted in figs.~\ref{f:slabav} a) and b); they mirror those simulated by \citet{siebesma1995evaluation}, which in turn reasonably match the apparent heat and moisture sources measured by \citet{nitta1974heat}.

These budgets quantify the effects of shallow cumulus convection on the slab-average thermodynamic structure. Up to the trade-inversion's base, convergence of moisture and heat fluxes moisten and heat the layer. Conversely, these fluxes moisten and cool the inversion layer, countering the drying and warming from the mean environment's subsidence. The imposed radiative source offers additional cooling throughout the layer. In spite of our intentions, and contrary to the models participating in \citet{siebesma2003large}, these processes do not quite balance, resulting in a negative $\overline{\theta_l}$ tendency and positive $\overline{q_t}$ tendency in the inversion layer: Our simulation is not a steady mean state. We will elaborate on this in section~\ref{s:dis}.

\subsection{The role of net condensation}

The appropriate thermodynamic quantity for analysing the capacity of the boundary layer to work against the subsiding environment is the buoyancy. It also bridges the apparent gap between the budgets of conserved variables discussed above, which do not refer to the presence of liquid water, and the role of net condensation, which is key to our analysis. 

We will interpret buoyancy through fluctuations in virtual potential temperature
$\theta_v$:

\begin{equation}
    \theta_v = \theta\left[1 + \left(\frac{R_v}{R_d} - 1\right)q_t - \frac{R_v}{R_d}q_l\right].
\end{equation}

Borrowing from \citet{stevens2007growth}, its fluctuations can to good approximation can be written as

\begin{equation}
    \theta_v' \approx a_1\theta_l' + a_2\overline{\theta_l}q_t' + a_3\overline{\theta_l}q_l'
    \label{e:buoy}
\end{equation}

\noindent with constants $a_1-a_3$ set to

\begin{subequations}
    \begin{equation}
        a_1 = \frac{\theta_{v_c}}{\theta_{l_c}}\frac{1}{1+\frac{q_{l_c}L_v}{c_pT_c}} \approx 1,  
    \end{equation}    
    \begin{equation}
        a_2 = 1-\frac{R_v}{R_d} \approx 0.608,
    \end{equation}
    \begin{equation}
        a_3 = a_1 \frac{\theta_{l_c} L_v}{\theta_c c_pT_c} - \frac{R_v}{R_d} \approx 7.
    \end{equation}
    \label{eq:thlvpconst}
\end{subequations}

In these relations, co-fluctuations among the thermodynamic variables are neglected, and $\theta_{v_c}$, $\theta_{l_c}$, $\theta_c$, $q_{l_c}$ and $T_c$ are taken to be representative cloud layer constants. $R_v$ is the ideal gas constants for dry moist air. If we reinterpret the spatial fluctuations in eq.~\ref{e:buoy} as changes occurring in time, we may also to good approximation write

\begin{equation}
    \frac{\partial \theta_{v}}{\partial t} \approx \frac{\partial \theta_{l}}{\partial t} + a_2\overline{\theta_l}\frac{\partial q_{t}}{\partial t} + a_3\overline{\theta_l}\frac{\partial q_{l}}{\partial t}.
    \label{e:buoyavtend}
\end{equation}


We will make two notes on eq.~\ref{e:buoyavtend} that prepare us for our subsequent discussion. First, we note that $a_3\overline{\theta_l}\frac{\partial q_{l}}{\partial t}$  is small in the slab-average \citep{betts1973non}, giving a powerful constraint that we return to in section~\ref{s:mec}~\ref{ss:thlvpm}. Consider eq.~\ref{e:slabav} with $\chi=q_l$, plotted in fig.~\ref{f:slabav} c). In our simulations, we do not impose large-scale horizontal transport of liquid water, while the effects of the prescribed subsidence on $\overline{q_l}$ are negligible. This allows us to simplify eq.~\ref{e:slabav} to

\begin{equation}
    \frac{\partial \overline{q_l}}{\partial t} = -\frac{1}{\rho_0}\frac{\partial}{\partial z}\left(\rho_0\overline{w'q_l'}\right) + \overline{\mathcal{C}} \approx 0,
    \label{e:qlavtend}
\end{equation}

\noindent where $\mathcal{C}$, the net condensation, is the only source in the absence of precipitation. Eqs.~\ref{e:buoyavtend} and~\ref{e:qlavtend} together explain how, in the conditionally unstable cloud layer, the positive net condensation is not stored as local potential energy ($\theta_v$), but is instead almost immediately exchanged into upwards transport; this is the balance shown in fig.~\ref{f:slabav} c). In the stable inversion layer, the transported liquid water entirely re-evaporates, resulting in a local convergence of $\overline{w'q_l'}$ that integrates to zero over the column. 

Applying the constraint eq.~\ref{e:qlavtend} to eq.~\ref{e:buoyavtend} brings us to our second note: Eq.~\ref{e:buoyavtend} reduces to an equation for another quantity that is conserved in non-precipitating convection, which \citet{grenier2001moist} call the liquid-water virtual potential temperature:

\begin{equation}
    \theta_{lv} = \theta_l + a_2\overline{\theta_l}q_t \equiv \theta_v - a_3\overline{\theta_l}q_l
    \label{e:thlv}
\end{equation}

Inserting eq.~\ref{e:slabav} with $\chi \in \left[\theta_l,q_t\right]$ in the slab-averaged eq.~\ref{e:buoyavtend} gives

\begin{equation}
	\frac{\partial \overline{\theta_v}}{\partial t} \approx \frac{\partial \overline{\theta_{lv}}}{\partial t} \propto -\frac{1}{\rho_0}\frac{\partial}{\partial z}\left(\rho_0\left( \overline{w'\theta_l'} + a_2\overline{\theta_l}\overline{w'q_t'} \right)\right),
\end{equation}

from which we may observe that the tendency of $\overline{\theta_v}$ is proportional not to the slab-averaged buoyancy flux $\overline{w'\theta_v'}$, but to fluxes of $\theta_{lv}$, which may be derived from eq.~\ref{e:buoy} by multiplication with $w'$:

\begin{equation}
    w'\theta_{lv}' = w'\theta_l' + a_2\overline{\theta_l}w'q_t' \equiv w'\theta_{v}' - a_3\overline{\theta_l}w'q_l'.
    \label{e:wthv}
\end{equation}

\begin{figure}
    \centering
    \includegraphics[width=0.5\textwidth]{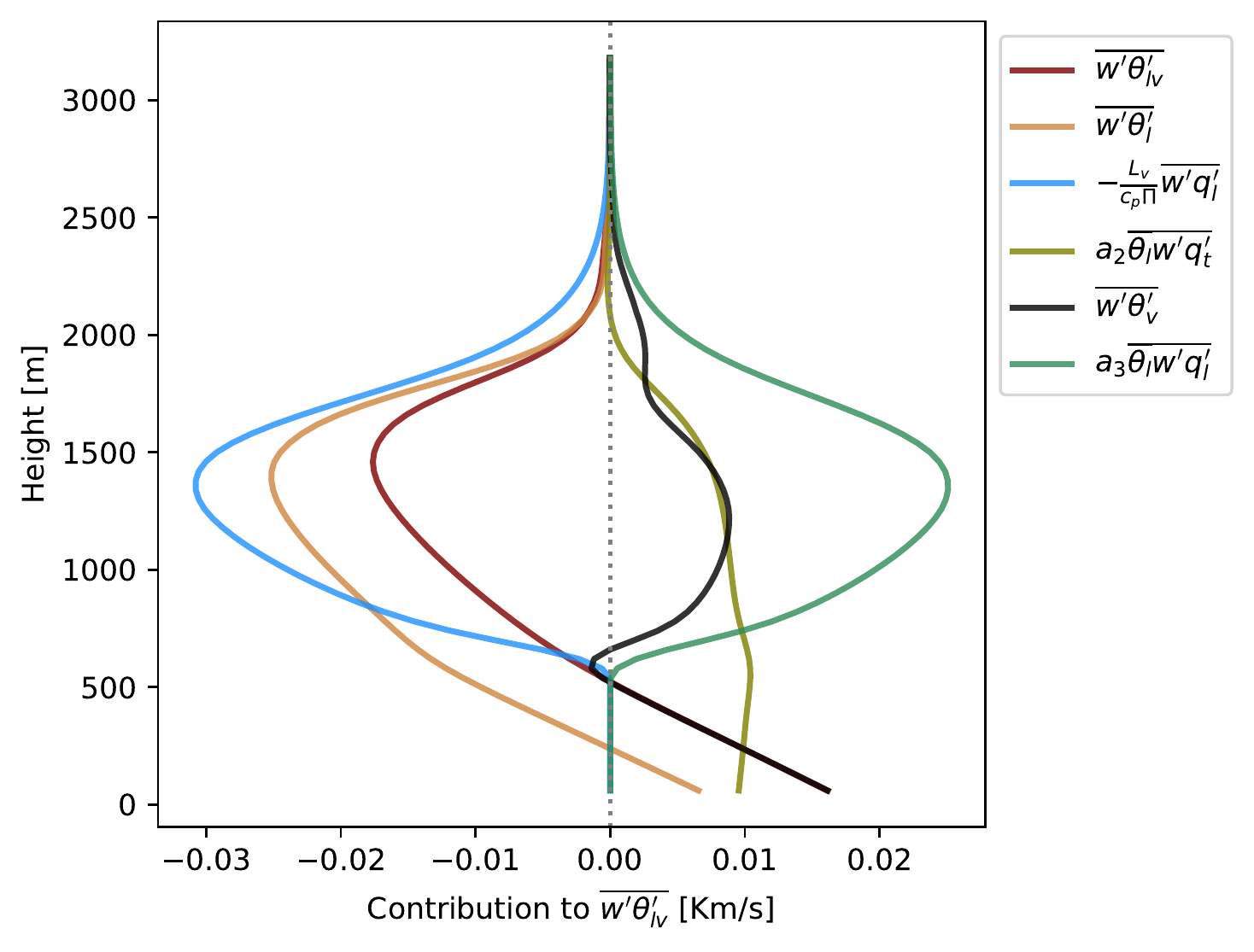}
    \caption{Contributions of the fluxes in eqs.~\ref{e:wthv} and~\ref{e:wthl} to $\overline{w'\theta_{lv}'}$, averaged over hours 6-16 of the simulation.}
    \label{f:wthlvpav}
\end{figure}

The slab-average of this flux, and its contributions, are plotted in fig.~\ref{f:wthlvpav}; this figure largely explains the effect of $\mathcal{C}$ on the layer. In one view of eq.~\ref{e:wthv}, the structure of $\overline{w'\theta_{lv}'}$ (maroon line) closely follows that of $\overline{w'\theta_{l}'}$ (yellow line). However, one may use eq.~\ref{e:thl} to write

\begin{equation}
    w'\theta_{l}' = w'\theta' - \frac{L_v}{c_p\Pi}w'q_l',
    \label{e:wthl}
\end{equation}

\noindent which led \citet{betts1975parametric} to recognise that it is mainly $L_v/(c_p\Pi)\overline{w'q_l'}$ (blue line) that is responsible for maintaining the large, downward $\overline{w'\theta_{l}'}$ in the cloud layer. The other view of eq.~\ref{e:wthv} follows \citet{stevens2007growth}: $w'\theta_{lv}'$ is the buoyancy flux, minus its contributions from liquid water fluxes. By definition, $w'\theta_v' = w'\theta_{lv}'$ in the subcloud layer. But in the cloud layer, fig.~\ref{f:wthlvpav} shows how the former is substantially outweighed by the latter: In energetic terms, latent heating and cooling translate more efficiently into vertical kinetic energy ($w'^2$) than into storage of local potential energy ($\theta_v'$). In both views, the structure of $\overline{w'\theta_{lv}'}$ is mainly supported by that of $\overline{w'q_l'}$. Since eq.~\ref{e:qlavtend} shows how $\mathcal{C}$ governs the net divergence of $\overline{w'q_l'}$ in the cloud layer and its convergence in the inversion layer, this discussion buttresses the classical picture of the cumulus-topped boundary layer that we drew in the introduction: Net condensation and the associated latent heating of the cloud layer, and the subsequent transport to, re-evaporation in and cooling of the inversion layer, balance the heating of the subsiding environment, and thus maintain the structure of $\overline{\theta_v}$ across the layer in general and across the trade inversion in particular.

\section{Summary of \citetalias{bretherton2017understanding}'s model for mesoscale fluctuations}
\label{s:mec}

We are now ready to summarise \citetalias{bretherton2017understanding}'s model for the development of mesoscale fluctuations. We will do so using only the classical theory outlined above, and a single assumption on the horizontal buoyancy field that also turns out to have similar consequences as we have already discussed. In section~\ref{s:add}, we will then move beyond \citetalias{bretherton2017understanding}'s theory, to a closed-form model of the instability and an analysis of its conditions.

\subsection{Definitions}

Following \citetalias{bretherton2017understanding}, we will frame length scale growth in our fields as an increase in magnitude of mesoscale fluctuations that develop over the slab-average. One can identify such mesoscale fluctuations in $\chi$ by partitioning $\chi'$, defined in eq.~\ref{e:fluc}, into a mesoscale component $\chi'_m$ and sub-mesoscale component $\chi'_s$, which gives:

\begin{equation}
    \chi = \overline{\chi} + \chi'_m + \chi'_s.
    \label{e:mesdef}
\end{equation}

\begin{figure}[t]
 \centering
 \noindent\includegraphics[width=0.66\textwidth]{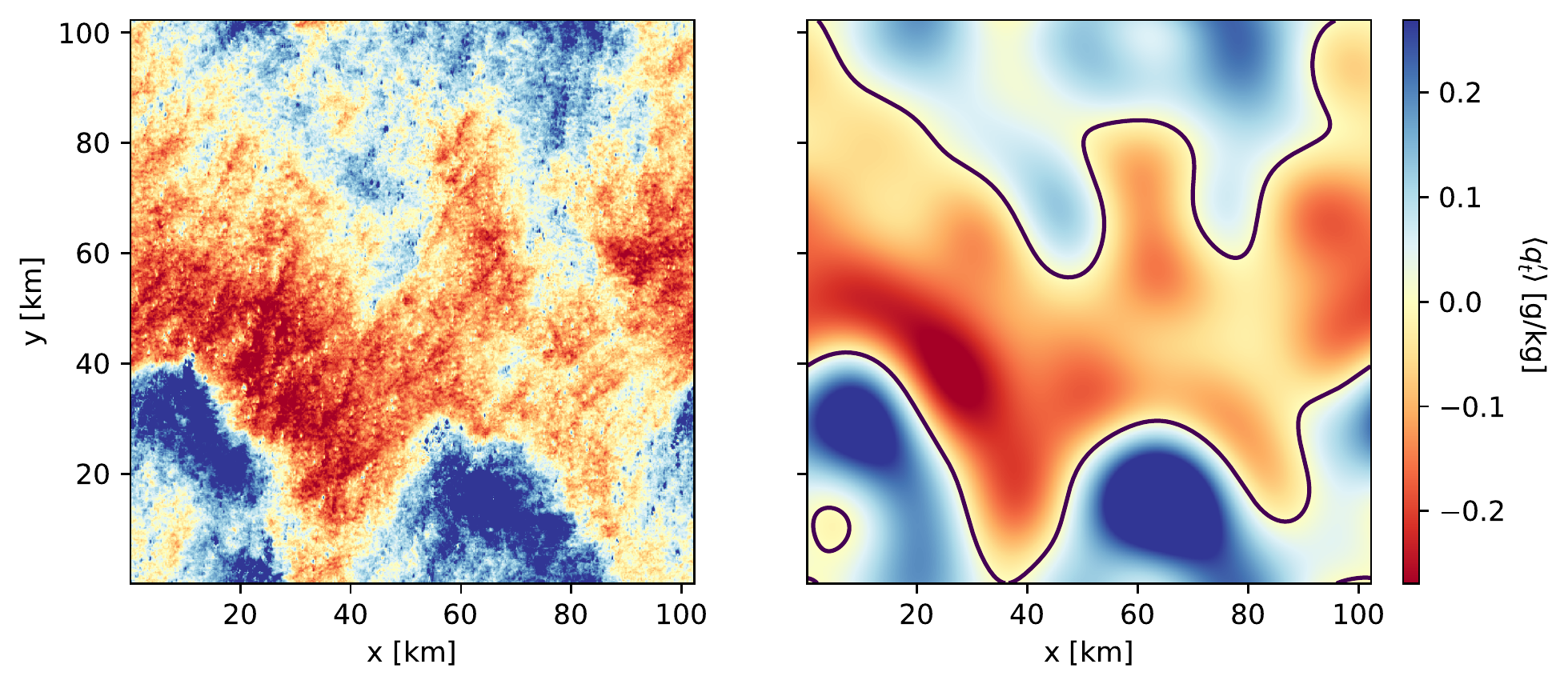}\\
 \caption{Fluctuations of column-averaged total specific humidity $\langle q_t \rangle$ (left), and its mesoscale-filtered component $\langle q_{t_m}'\rangle$, overlaid with a contour separating mesoscale regions that are moister and drier than the spatial mean (right) after 24 hours.}\label{f:twpfluct}
\end{figure}

\citetalias{bretherton2017understanding} scale-partition their variables by extracting horizontal averages over blocks of 16 km $\times$ 16 km. Here, we conduct the decomposition with a spectral low-pass filter at the horizontal wavenumber that corresponds to scales of 12.5 km. As an example, consider fig.~\ref{f:twpfluct}; our spectral filter extracts the field shown in the right panel from that shown in the left panel. Of course, any choice of method and scale for this separation is somewhat arbitrary. Yet, since its primary objective is to distinguish mesoscale fluctuations from fluctuations that occur on the scale of a typical cumulus cloud, we expect that any consistently performed scale separation at a scale that is larger than this typical cumulus scale (around 1 km), but sufficiently smaller than our finite domain size (100 km), would suffice to illustrate what we intend to show.


In the following, special attention will be paid to ``moist, mesoscale regions''. To define such regions, we use the density-weighted vertical average

\begin{equation}
    \langle \chi \rangle = \frac{\int_0^{z_{\infty}}\rho_0\chi dz}{ \int_0^{z_{\infty}}\rho_0 dz},
    \label{e:vint}
\end{equation}

\noindent where $z_{\infty}$ refers to the domain top at 10 km. From this follows a definition of the column-averaged, or bulk moisture $\langle q_t \rangle$. We will take moist mesoscale regions to be horizontal coordinates where $\langle q_{t_m}' \rangle>0$, and dry mesoscale regions where $\langle q_{t_m}' \rangle<0$. The black contour line in the right panel of fig.~\ref{f:twpfluct} gives a visual impression of this delineation.

With these definitions, we formulate a budget for $\chi_m'$, by subtracting eq.~\ref{e:slabav} from eq.~\ref{e:localbudgrew}, making several assumptions that make the equation consistent with our LES, mesoscale-filtering the result, and rearranging the terms (see appendix A for the derivation's details):


\begin{equation}
    \frac{\partial \chi_m'}{\partial t} =
    -w_m'\Gamma_{\chi}
    - \frac{1}{\rho_0}\frac{\partial}{\partial z}\left( \rho_0 F_{\chi'_m} \right)
    - \frac{\partial}{\partial x_{jh}}\left(u_{jh} \chi'\right)_m
    - \overline{w}\frac{\partial \chi_m'}{\partial z}
    + S_{\chi_m}'.
    \label{e:chipf}
\end{equation}

In this relation, the mean vertical gradient $\partial \overline{\chi}/\partial z = \Gamma_{\chi}$ and $F_{\chi_m'}$ is the anomalous mesoscale vertical flux of $\chi'$ away from the slab average

\begin{equation}
     F_{\chi_m'} = \left(w'\chi'\right)_m - \overline{w'\chi'}.
     \label{e:fchi}
\end{equation}

In spite of the number of steps taken to derive it, we draw attention to eq.~\ref{e:chipf}'s similarity to the slab-averaged budget, eq.~\ref{e:slabav}. It features the horizontal and vertical convergence of $\chi_{m}'$ (second and third terms on the right-hand side of eq.~\ref{e:chipf}), and the mesoscale anomalous effect of subsidence (fourth term) and sources (fifth term). The most important difference with eq.~\ref{e:slabav} is the first term, which describes transport along the mean gradient of $\chi$ with mesoscale vertical velocity fluctuations. This term will prove to be central.


Many of our results will show averages of eq.~\ref{e:chipf} over moist and dry mesoscale regions, which, because such regions are not entirely stationary, introduces two nuances (see appendix A c). First, it allows us to remove the effects of transport of the mesoscale fluctuations with $\overline{u_{jh}}$, and second, it introduces a divergence term that describes the net expansion of the regions with velocity $u_{jh}^e$. Denoting averages over moist or dry regions by $\widetilde{\cdot}$, this finally gives:

\begin{multline}
    \underbrace{\frac{\partial \widetilde{\chi_m'}}{\partial t}}_{\text{Tendency}} =
    \underbrace{\widetilde{-w_m'}\Gamma_{\chi}}_{\text{Gradient production}}
    - 
    \underbrace{\frac{\partial}{\partial x_{j_h}}\widetilde{\left(u_{jh}' \chi'\right)_m}}_{\text{Horizontal transport}} 
    + 
    \underbrace{\frac{\partial}{\partial x_{jh}}\widetilde{\left(u_{jh}^e \chi'\right)_m}}_{\text{Expansion}}
    \\
    - 
    \underbrace{\frac{1}{\rho_0}\frac{\partial}{\partial z}\left( \rho_0 \widetilde{F_{\chi'_m}} \right)}_{\text{Vertical transport}}
    -
    \underbrace{\overline{w}\frac{\partial \widetilde{\chi_m'}}{\partial z}}_{\text{Subsidence}}
    + 
    \underbrace{\widetilde{S_{\chi_m}'}}_{\text{Source}}
    \label{e:chipfav}
\end{multline}

To keep the text uncluttered, we will only discuss explicitly regions where $\langle q_{t_m}' \rangle>0$, since observations pertaining to such regions are generally the opposite in dry, mesoscale regions, albeit of different magnitude. To give the reader an impression of this asymmetry, the figures will generally present both moist and dry profiles.

\subsection{A sketch of the instability}
\label{ss:mes}

\begin{figure}[t]
 \centering
 \noindent\includegraphics[width=\textwidth]{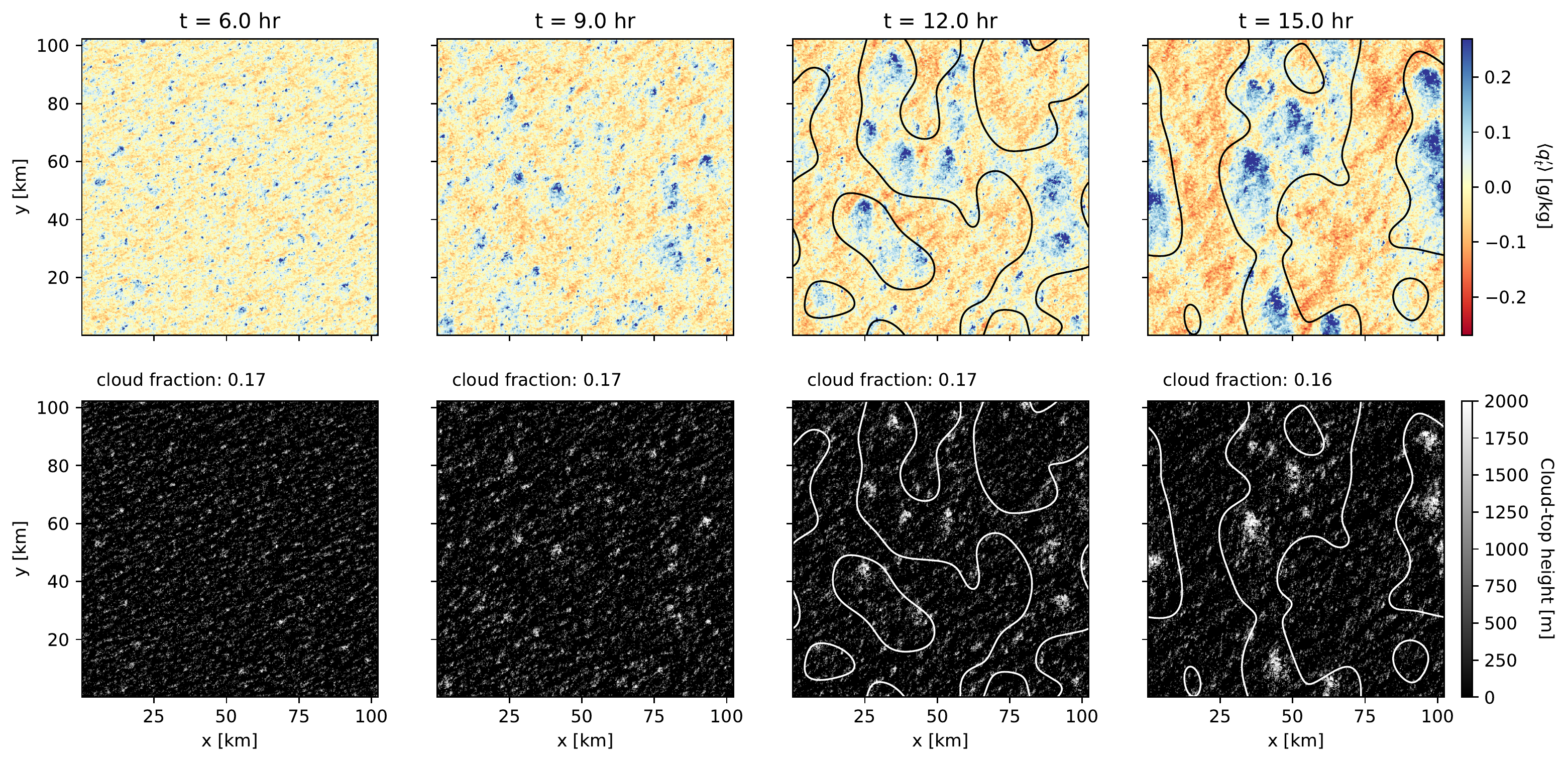}\\
 \caption{Time evolution (left to right) of $\langle q_{t_m}' \rangle$ (top row) and cloud top height (bottom row), overlaid by contours that separate moist and dry mesoscale regions at 12 and 15 hours, and annotated with cloud fraction.}\label{f:twp_cld_evo}
\end{figure}

\begin{figure}[t]
 \centering
 \noindent\includegraphics[width=\textwidth]{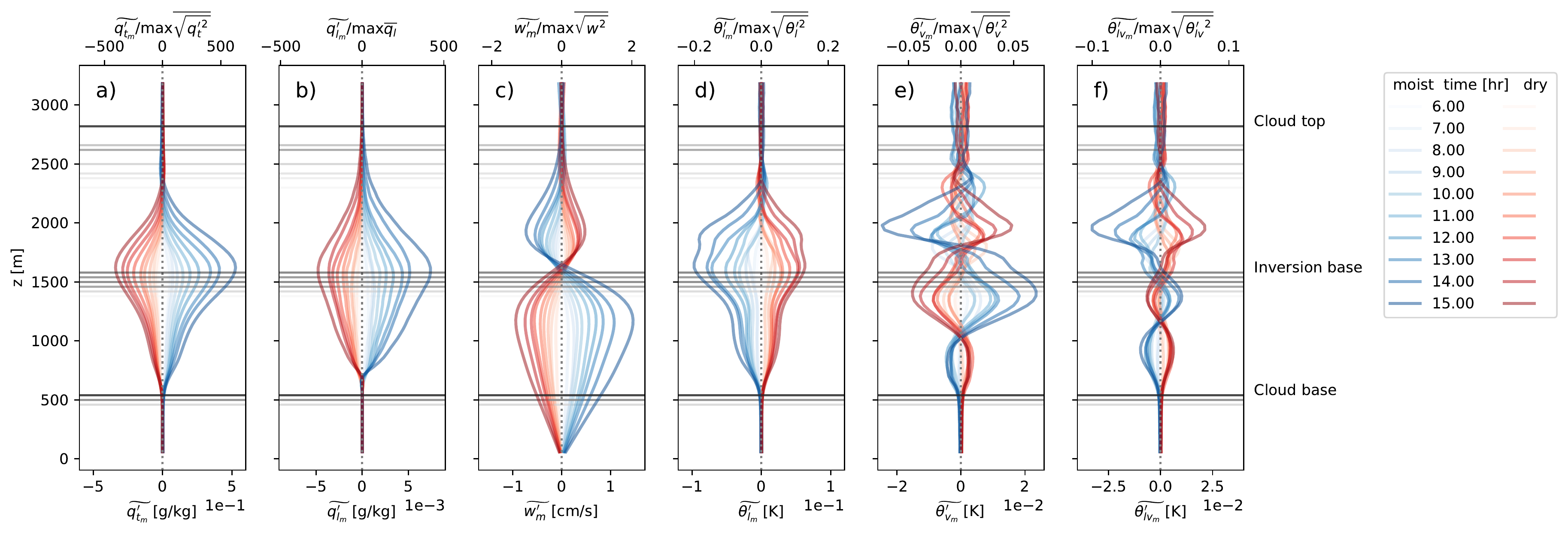}
 \caption{Time evolution of mesoscale fluctuations, averaged over moist and dry mesoscale regions, for total specific humidity $q_t$ (a), liquid-water specific humidity $q_l$ (b), vertical velocity $w$ (c), liquid-water potential temperature $\theta_l$ (d), virtual potential temperature $\theta_v$ (e) and liquid-water virtual potential temperature $\theta_{lv}$ (f). Upper axes indicate the maximum  of these fluctuations relative to the maximum root-mean-square fluctuation in each quantity at the last, plotted time.}\label{f:vars_meso_evo}
\end{figure}

The top row of fig.~\ref{f:twp_cld_evo} shows how small disturbances in $\langle q_t'\rangle$ grow into significant mesoscale fluctuations over an eight hour time window. The figure's bottom row identifies growing clusters of shallow cumulus clouds that develop on top of these mesoscale regions, becoming more vigorous and reaching deeper into the inversion as they grow. Fig.~\ref{f:vars_meso_evo} a) and b) quantifies this in the form of vertical profiles. Since the fluctuations in the temperature variables discussed in section~\ref{s:ana} remain small with respect to their root-mean-square (see fig.~\ref{f:vars_meso_evo} d-f), this suggests that to understand the length scale growth of our clouds, we must understand what drives the formation of $\langle q_{t_m}' \rangle$.

\begin{figure}[t]
 \centering
 \noindent\includegraphics[width=\textwidth]{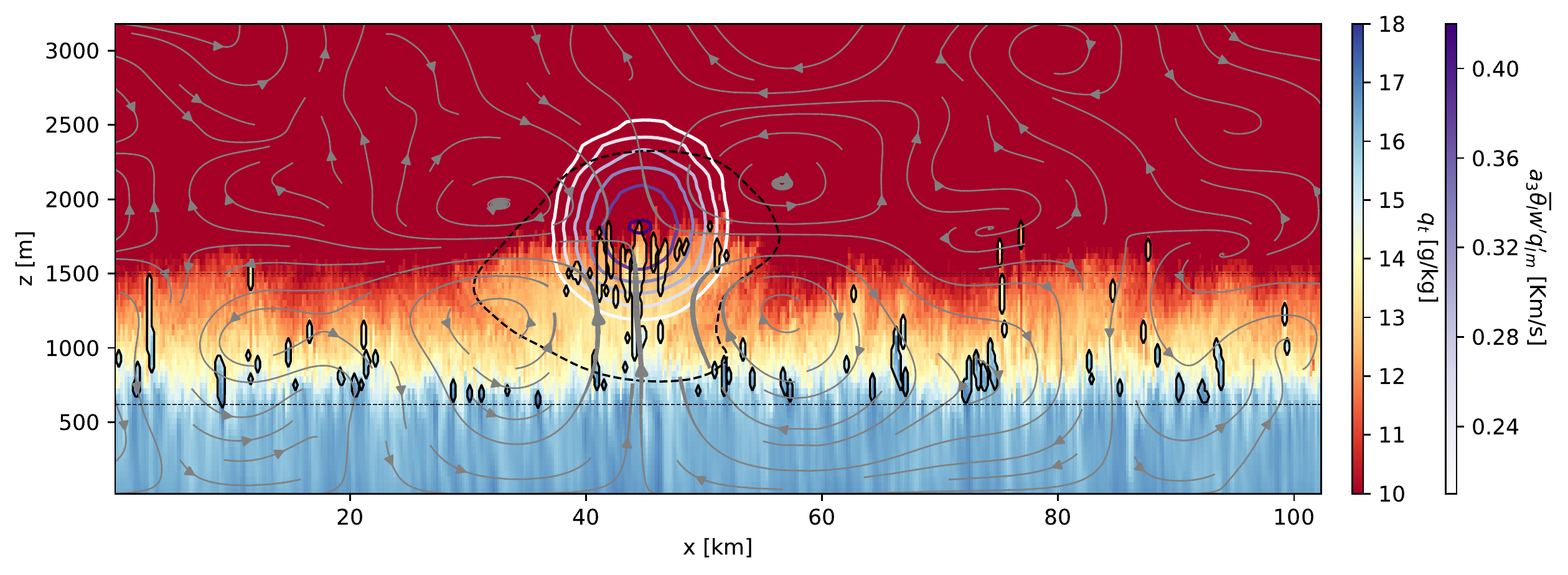}\\
 \caption{Cross-section over an example y-z plane of our simulation at 16 hours, coloured by filled contours of $q_t$ (red to blue) and overlaid by contour lines of i) $q_{t_m}'>0$ (black, dashed), ii) clouds (black, unbroken) and iii) $\left(w'q_l'\right)_m$ (white to blue; it does not coincide with the clouds because it is filtered at slab level). Also overlaid are streamlines of the mesoscale-filtered, in-plane velocity fluctuations (defined by $\left[u_m',w_m'\right]$), whose line thickness is weighted by this velocity's local magnitude. Horizontal, dashed lines represent cloud base and inversion base.}\label{f:concept}
\end{figure}

Fig.~\ref{f:concept} offers a sketch of the explanation. Over the vertical dimension, clouds (black contour lines) develop favourably on top of a patch of $q_{t_m}'>0$ (black, dashed contour) in the upper cloud layer. The $q_{t_m}'$ structure is produced by a mesoscale circulation of approximately 1 cm/s (overlaid streamlines), which converges in the sub-cloud layer beneath the structure, transports moisture upwards along the negative, slab-mean vertical moisture gradient, and detrains it laterally near the inversion base around 1500m, where the mesoscale vertical velocity $w_m'$ becomes negative. Fig.~\ref{f:vars_meso_evo}, which shows the temporal development of mesoscale flucutations in $q_t$, $q_l$ and $w_m'$, averaged over moist and dry mesoscale regions, quantifies these statements. The reader will recognise that we have here merely described the action of the first term in eq.~\ref{e:chipf} and eq.~\ref{e:chipfav}; we will make this connection explicit in section~\ref{s:mec}\ref{ss:qtpm}.

The mesoscale circulations themselves arise from corresponding mesoscale variations in the classical theory of the slab-averaged layer that we have discussed in section~\ref{s:ana}, supplemented by a single, well-known assumption from mesoscale tropical meteorology, namely that horizontal fluctuations in density remain small. We observe the resulting ``weak temperature gradients'' in the profiles of mesoscale buoyancy fluctuations $\theta_{v_m}'$, plotted in fig.~\ref{f:vars_meso_evo} e), which do not differ appreciably between moist and dry mesoscale regions. This allows the circulations to develop directly from mesoscale fluctuations in condensational heating in the cloud layer, and evaporative cooling in the inversion layer; we show this in section~\ref{s:mec}\ref{ss:thlvpm}. With reference to our discussion in section~\ref{s:ana}, such heating fluctuations are anticipated by the mesoscale-filtered vertical flux of liquid water in fig.~\ref{f:concept} (white-to-blue contours). 

The mesoscale condensation anomalies again favour regions with positive mesoscale moisture fluctuations, which control the mesoscale relative humidity fluctuations when the (potential) temperature fluctuations, shown in fig.~\ref{f:vars_meso_evo} d)-f) are small. In all, \citetalias{bretherton2017understanding} then identify a self-reinforcing feedback: Mesoscale fluctuations in condensation and evaporation in cumulus clouds give rise to mesoscale circulations, which in turn enhance mesoscale moisture fluctuations, on top of which stronger mesoscale fluctuations in condensation and evaporation develop. 

\subsection{Mesoscale moisture fluctuations develop from mesoscale circulations}
\label{ss:qtpm}

\begin{figure}[t]
    \centering
    \includegraphics[width=0.9\textwidth]{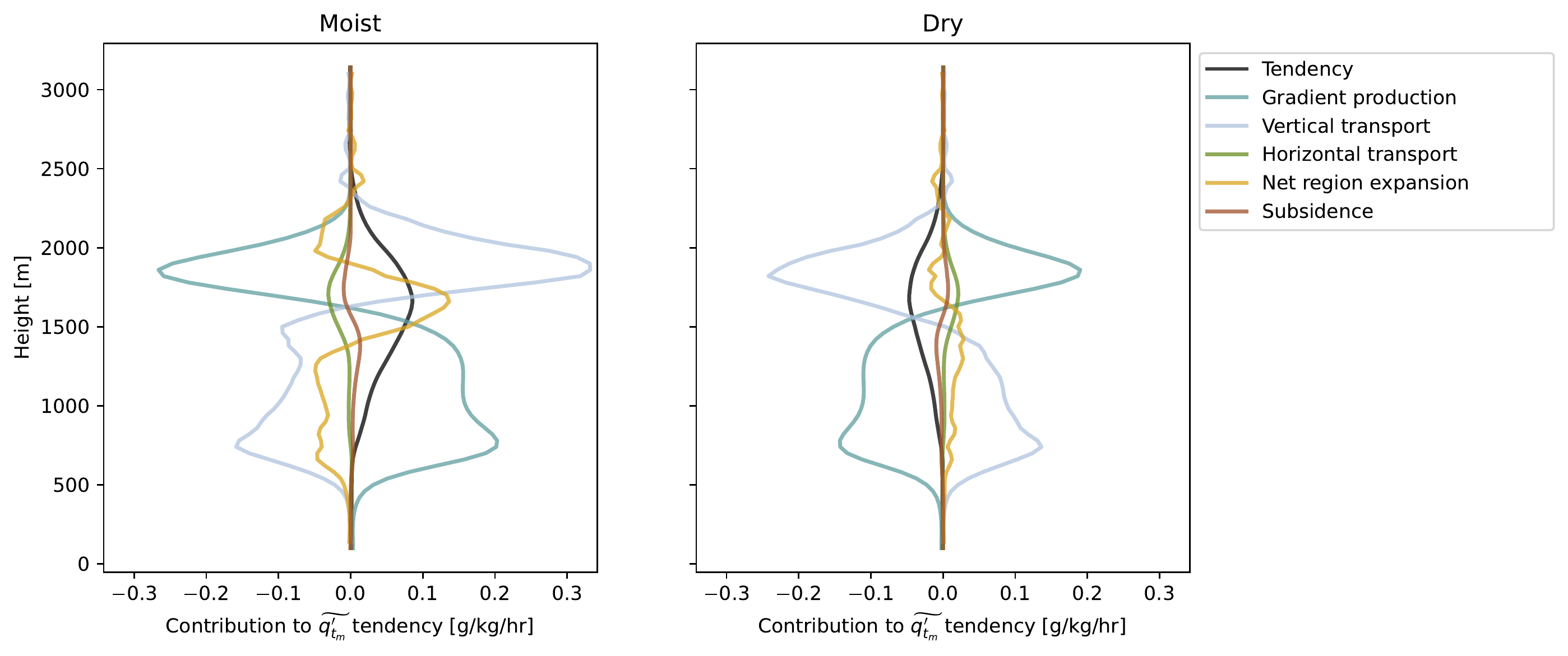}
    \caption{Vertical profiles of the terms in the $q_{t_m}'$ budget averaged over moist and dry mesoscale regions (eq.~\ref{e:chipfav}), and over 10-16 hr.}
    \label{f:qtpf_budget}
\end{figure}

Fig.~\ref{f:qtpf_budget} shows the terms in eq.~\ref{e:chipfav} with $\chi=q_t$. It identifies the main reason for the rise of $q_{t_m}'>0$ in the moist cloud layer to be the production of $q_{t_m}'$ by vertical, mesoscale transport along the mean, negative moisture gradient $\Gamma_{q_t}$. We will call this ``gradient production'', in the spirit of variance-budget studies \cite[e.g.][]{de2004large,heinze2015second,anurose2020understanding}, which show that this term, when scaled with the moisture flucutation itself, is the main driver of moisture variance in cloud-topped boundary layers. Our gradient production essentially quantifies a similar process as variance production, with the exception of tracing it in the mesoscales specifically, and with the advantage of being more easily visualised. The term follows directly from the $w_m'$ profiles plotted in fig.~\ref{f:vars_meso_evo} c). These are increasingly positive in the moist cloud layer, and increasingly negative in the moist inversion layer, and thus accelerate the gradient production in time.

Fig.~\ref{f:qtpf_budget} also shows that gradient production of $q_{t_m}'$ is, to a large extent, balanced by the convergence of vertical moisture fluxes. These fluxes transport the positive moisture fluctuation that is produced in the moist cloud layer into the overlying inversion. Since the term's vertical integral is zero, it does not add or remove $\langle q_{t_m}' \rangle$ from a column; it just translates the vertical structure of $-w_m'\Gamma_{q_t}$ into profiles of its tendency (black line in fig.~\ref{f:qtpf_budget}). It is therefore also mainly responsible for situating the peak of the mesoscale fluctuations in cloudiness ($q_{l_m}'$, see fig.~\ref{f:vars_meso_evo} b) in the inversion layer.

Horizontal transport enters the budget through i) the mesoscale horizontal moisture fluxes from moist to dry mesoscale regions and ii) the net region expansion with $u_{jh}^e$. In the cloud layer, the region expansion (yellow line) is negative, i.e. the mesoscale circulation's positive cloud-layer moisture convergence acts to concentrate the mesoscale moisture fluctuation. In the inversion, the work done in broadening the inversion-layer moisture fluctuation by the mesoscale, horizontal, outflows shown in fig.~\ref{f:concept} makes the term positive. This effect generally outweighs the drying (olive line) from transport across the region boundary with the mesoscale flow and through turbulent mixing of $q_{t_m}'$ down the horizontal moisture gradient, which draws $q_{t_m}'$ from the mesoscales.

The subsidence term is only a small direct contributor to the budget; as we have seen, its primary role is in setting the slab-mean environment in which the moisture fluctuations can develop. The budget has no further sources, i.e. in the absence of precipitation, $S_{{q_t}_m}'=0$.

\begin{figure}[t]
    \centering
    \includegraphics[width=\textwidth]{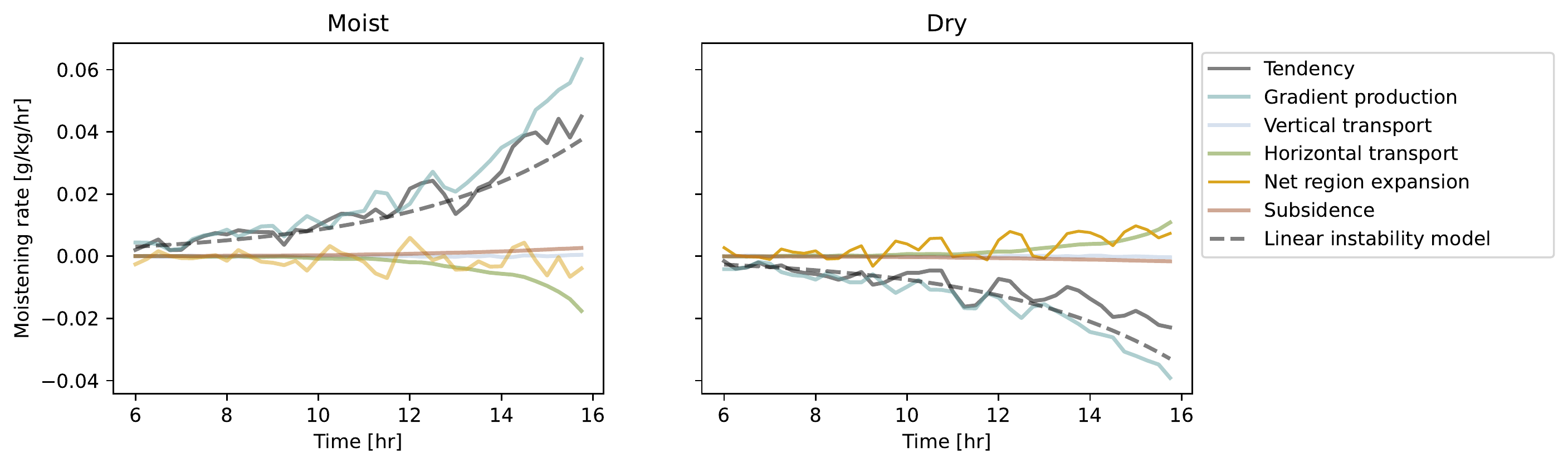}
    \caption{Time-evolution of the $\langle q_{t_m}' \rangle$ budget (eq.~\ref{e:vint} applied to eq.~\ref{e:chipfav}), averaged over moist and dry mesoscale regions, between 6-16 hr, in the spirit of fig.13 of \citet{bretherton2017understanding}. The dashed line plots the linear instability eq.~\ref{e:instabmodel}.}
    \label{f:qtpf_budget_int}
\end{figure}

The relative importance of the gradient production and horizontal advection of $q_{t_m}'$ to the development of moist and dry mesoscale regions is adequately captured by vertically averaging eq.~\ref{e:chipfav} with $\chi=q_t$ using eq.~\ref{e:vint}, which gives a budget for $\widetilde{\langle q_{t_m}' \rangle}$; the time-evolution of this budget is plotted in fig.~\ref{f:qtpf_budget_int}. It shows that the column-averaged mesoscale moistening rate increases roughly exponentially in moist areas, and that it is initially well-approximated by the gradient production. Only once significant mesoscale moist patches have formed (see fig.~\ref{f:twp_cld_evo} at 12 and 15 hours) do horizontal fluxes begin substantially opposing it. This suggests that $q_{t_m}'$ development is initially production-driven, while the horizontal structure and mature development of the fluctuations also depend on the efficiency with which horizontal transport can redistribute them.

The net expansion's column average is small, but slightly negative in moist areas, i.e. not only are large, moist areas becoming moister, they are also becoming slightly smaller. We will briefly discuss this clustering tendency and its significance in section~\ref{s:dis}. 

\subsection{Mesoscale circulations develop from anomalous latent heating in clouds}
\label{ss:thlvpm}

\subsubsection{Weak temperature gradients}

To understand why moist mesoscale regions grow, we must deduce the source of $w_m'$ in the gradient production of $q_{t_m}'$. In other words: Why does the mesoscale circulation shown in fig.~\ref{f:concept} develop? \citetalias{bretherton2017understanding} argue that this is best understood through a Weak Temperature Gradient (WTG) framework \citep[e.g.][]{held1985large,sobel2001weak}, which has proven useful in explaining self-organised, circulation-driven scale growth in the moisture fields of tropical atmospheres in radiative-convective equilibrium \citep[e.g.][]{emanuel2014radiative,chikira2014eastward,beucler2018linear,ahmed2019explaining}.

While deep convective clouds are the most spectacular example, any sort of convection in a stably stratified fluid generates density fluctuations which gravity waves continually redistribute horizontally, also the shallow cumuli under consideration here. Since these waves travel at a characteristic speed much higher than that with which advection can transport mixed scalars such as moisture, they may prevent buoyancy fluctuations from accumulating over the time scale with which $q_{t_m}'$ grows, i.e. the mesoscale buoyancy fluctuations remain small. A WTG interpretation of the governing equations then allows using the mesoscale buoyancy budget to diagnose $w_m'$.

To understand this, we first mesoscale filter our definition for buoyancy (eq.~\ref{e:buoy}):

\begin{equation}
    \theta_{v_m}' = \theta_{l_m}' + a_2\overline{\theta_l}q_{t_m}' + a_3\overline{\theta_l}q_{l_m}'
    \label{e:buoyf}
\end{equation}

From eq.~\ref{e:buoyf}, it is not immediately obvious that $\theta_{v_m}'$ should be small. For instance, if $\theta_{l_m}'\approx 0$, $\theta_{v_m}' \approx \overline{\theta_l}\left(a_2q_{t_m}' + a_3q_{l_m}'\right)$. So, upon inspecting fig.~\ref{f:vars_meso_evo} a) and b), one may expect to find mesoscale buoyancy fluctuations that correlate to the moisture and liquid water fluctuations. This turns out to be a very good approximation in layers of relatively continuous cloud cover, such as closed cell convection \citep{de2004large,de2008effect}, in which thermals hardly penetrate the stable layer above the mixed layer and thus give gravity waves much less of a chance to redistribute their buoyancy. In such situations, significant $\theta_{v_m}'$ are observed to develop in the cloud layer, which may contribute directly to the development of $w_m'$ through the mesoscale-filtered vertical momentum equation.

\begin{figure}[t]
    \centering
    \includegraphics[width=0.9\textwidth]{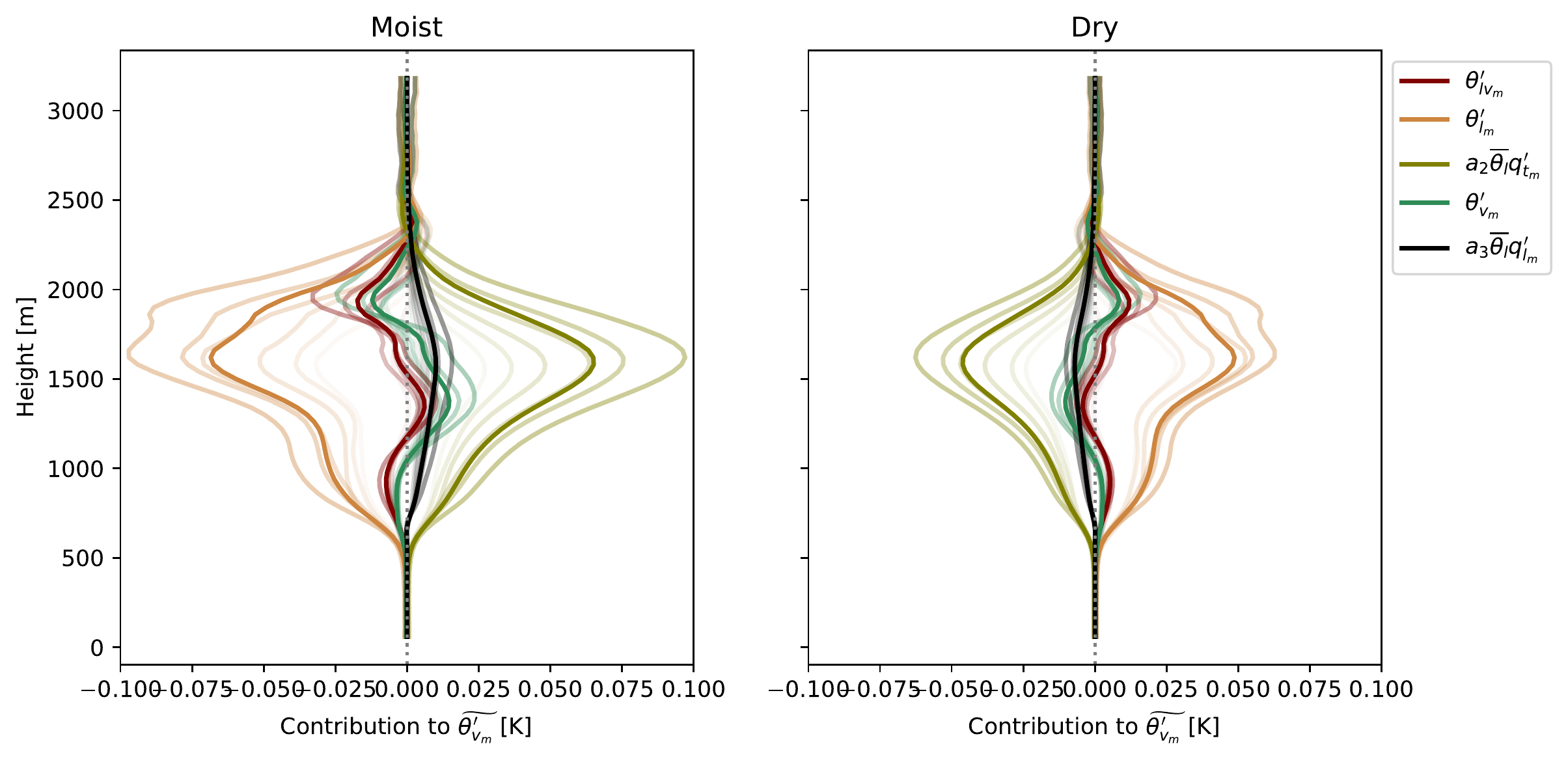}
    \caption{Vertical profiles of $\theta_{v_m}'$ (eq.~\ref{e:buoy}) and $\theta_{lv_m}'$ (eq.~\ref{e:thlv}), and their contributions from $\theta_{l_m}'$, $q_{t_m}'$ and $q_{l_m}'$, averaged over moist and dry mesoscale regions (left, right), over 10-16 hr (dark lines) and over every hour between 10-16 hr (transparent lines, in order of increasing opacity).}
    \label{f:thvp}
\end{figure}

However, our broken cumulus layer demands a different view: Fig.~\ref{f:thvp}, which plots the contributions to eq.~\ref{e:buoyf} (opaque lines) and their time evolution (increasingly dark, transparent lines), indicates $\widetilde{\theta_{l_m}'} \neq 0$ and $a_2\overline{\theta_l}\widetilde{q_{t_m}'}\gg a_3\overline{\theta_l}\widetilde{q_{l_m}'}$. Instead of remaining small, $\theta_{l_m}'$ becomes increasingly negative in moist mesoscale regions, while $a_3\overline{\theta_l}\widetilde{q_{l_m}'}$ remains almost negligible. The result is an approximate balance between the $\theta_{l_m}'$ and $q_{t_m}'$ contributions in eq.~\ref{e:buoyf}, and a comparatively small $\widetilde{\theta_{v_m}'}$ (black line in fig.~\ref{f:thvp}) with little temporal development. Hence, if we differentiate eq.~\ref{e:buoyf} to time, we may write the mesoscale-fluctuation equivalent of eq.~\ref{e:buoyavtend}, and recognise that it is stationary, akin to its slab-averaged counterpart being steady to maintain a stable trade-inversion:

\begin{equation}
    \frac{\partial \theta_{v_m}'}{\partial t} \approx \frac{\partial \theta_{l_m}'}{\partial t} + a_2\overline{\theta_l}\frac{\partial q_{t_m}'}{\partial t} + a_3\overline{\theta_l}\frac{\partial q_{l_m}'}{\partial t} \approx 0.
    \label{e:buoytend}
\end{equation}

Eq.~\ref{e:buoytend} is our statement of the WTG approximation. Before using it, we pause to employ the observation that the temporal development of $a_3\overline{\theta_l}\widetilde{q_{l_m}'}$ does not seem to appreciably influence $\widetilde{\theta_{v_m}'}$ (fig.~\ref{f:thvp}). Indeed, we may approximate eq.~\ref{e:chipf} for $q_l$ as a slightly more stringent version of eq.~\ref{e:qlavtend}:

\begin{equation}
    \frac{\partial q_{l_m}'}{\partial t} \approx -\frac{\partial}{\partial z}\left(F_{q_{l_m}'}\right) + \mathcal{C}_m' \approx 0.
    \label{e:qlpf}
\end{equation}

\begin{figure}[t]
    \centering
    \includegraphics[width=0.9\textwidth]{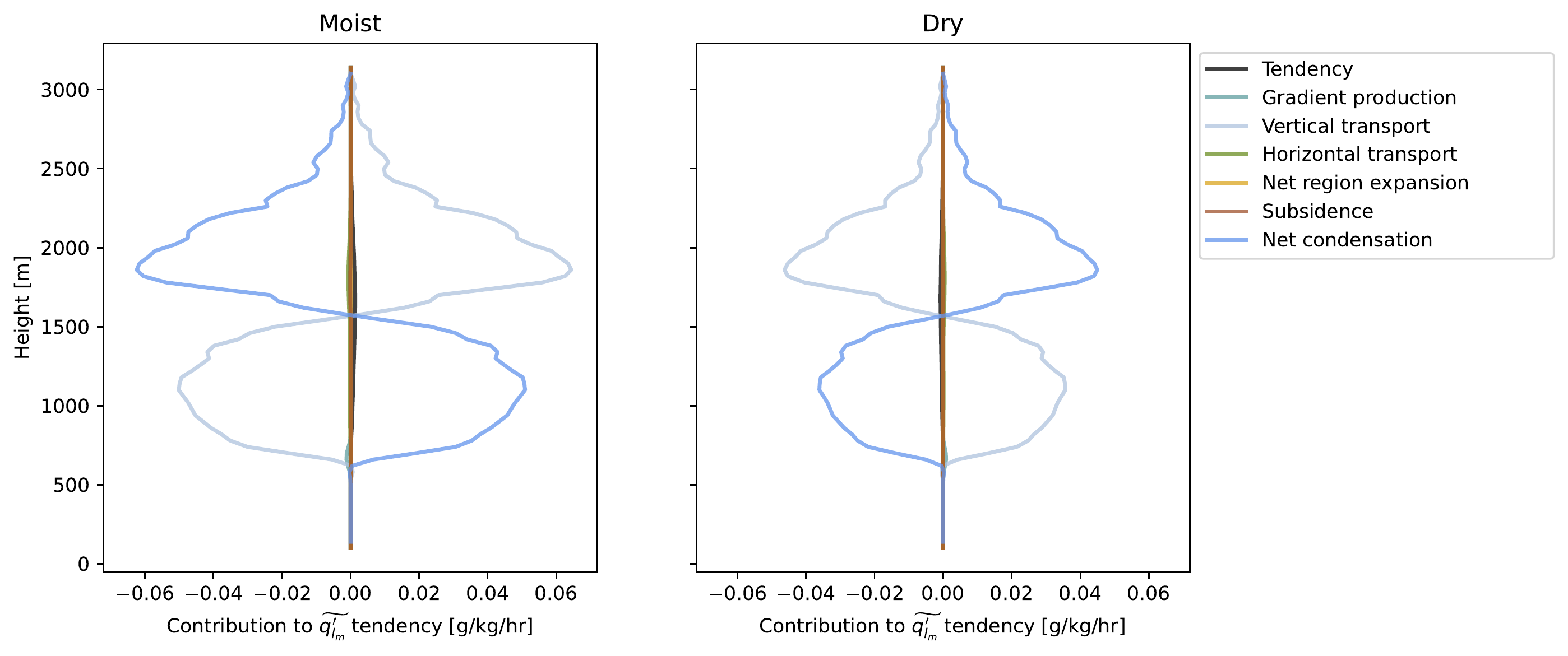}
    \caption{Vertical profiles of the terms in the $q_{l_m}'$ budget averaged over moist and dry mesoscale regions (eq.~\ref{e:chipfav}, which approximately reduces to eq.~\ref{e:qlpf}), and over 10-16 hr.}
    \label{f:qlpf_budget}
\end{figure}

Hence, in another parallel of the slab-mean theory, mesoscale anomalies in the rate of net condensation $\mathcal{C}$ are approximately balanced by anomalies in their vertical transport, represented by the divergence of $F_{q_{l_m}'}$, making storage of $q_{l_m}'$ passive. These assumptions are confirmed by fig.~\ref{f:qlpf_budget}.

The upshot of this discussion is that even though we in section~\ref{s:ana} posed budgets for $\theta_v$ to analyse the stability of the trade-wind layer's slab-averaged structure to vertical growth, while we here pose it to analyse the growth of horizontal fluctuations, the consequences of applying eq.~\ref{e:qlpf} are similar: Eq.~\ref{e:buoyf} again reduces to a budget for $\theta_{lv}$, only here for its mesoscale fluctuations. Because eq.~\ref{e:qlpf} holds, $\theta_{lv_m}'$ satisfies the WTG approximation as well as $\theta_{v_m}'$. In adopting it, we again follow \citetalias{bretherton2017understanding}, who perform their analysis in terms of liquid virtual static energy, the energy-equivalent to $\theta_{lv}$.

\subsubsection{Modelling $w'_m$}

The budget for $\theta_{lv_m}'$ follows from inserting the (scaled) tendencies of $\theta_{l_m}'$ and $q_{t_m}'$, as written in eq.~\ref{e:buoytend}, into eq.~\ref{e:chipf}. Subsuming transport and (diabatic) sources of $\theta_{lv_m}'$ under $S_{\theta_{lv_m}}'$ then gives the WTG formulation on a commonly written form:

\begin{equation}
    \frac{\partial \theta_{lv_m}'}{\partial t} =
    -w_m'\frac{\partial\overline{\theta_{lv}}}{\partial z}
    + S_{\theta_{{lv}_m}}' \approx 0.
    \label{e:ddtthvpf}
\end{equation}

\begin{figure}[t]
    \centering
    \includegraphics[width=0.9\textwidth]{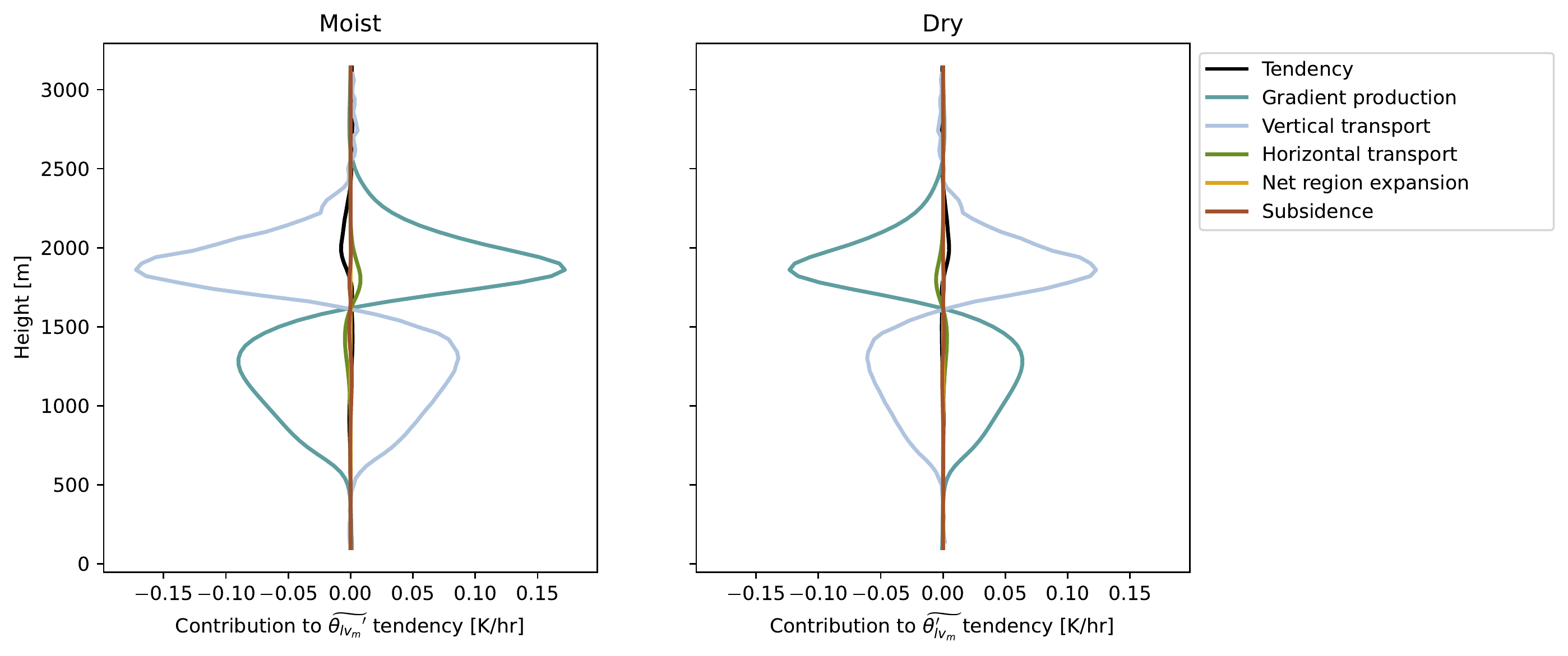}
    \caption{Vertical profiles of the terms in the $\theta_{lv_m}'$ budget (eq.~\ref{e:chipfav}, which approximately reduces to eq.~\ref{e:ddtthvpf}), averaged over moist and dry mesoscale regions, and over 10-16 hr.}
    \label{f:thlvpf_budget}
\end{figure}

Fig.~\ref{f:thlvpf_budget} plots all terms in this budget. It reveals that the gradient production is primarily balanced by anomalous, vertical mesoscale convergence of $\theta_{lv}'$ fluxes ($F_{{\theta_{lv}'}_m}$); both terms are at least an order of magnitude larger than $\theta_{lv_m}'$'s tendency, horizontal advection and subsidence heating. Hence, the figure invites us to rewrite eq.~\ref{e:ddtthvpf} as a diagnostic equation for $w_m'$, which can be understood as the vertical velocity needed to move air parcels heated by $S_{\theta_{lv_m}}'$ quasi-statically to their level of neutral buoyancy under a stable stratification \citep{klein2010scale}. In our non-precipitating simulations with homogenoud, imposed radiation, the anomalous heat source $S_{\theta_{lv_m}'}$ is reduced to the vertical convergence of $F_{{\theta_{lv}'}_m}$. Hence, the model reads:

\begin{equation}
    w_m' \approx - \frac{1}{\rho_0}\frac{\partial}{\partial z}\left( \rho_0 F_{{\theta_{lv}'}_m} \right) \frac{1}{\Gamma_{\theta_{lv}}}.
    \label{e:wwtg}
\end{equation}

\begin{figure}[t]
    \centering
    \includegraphics[width=0.9\textwidth]{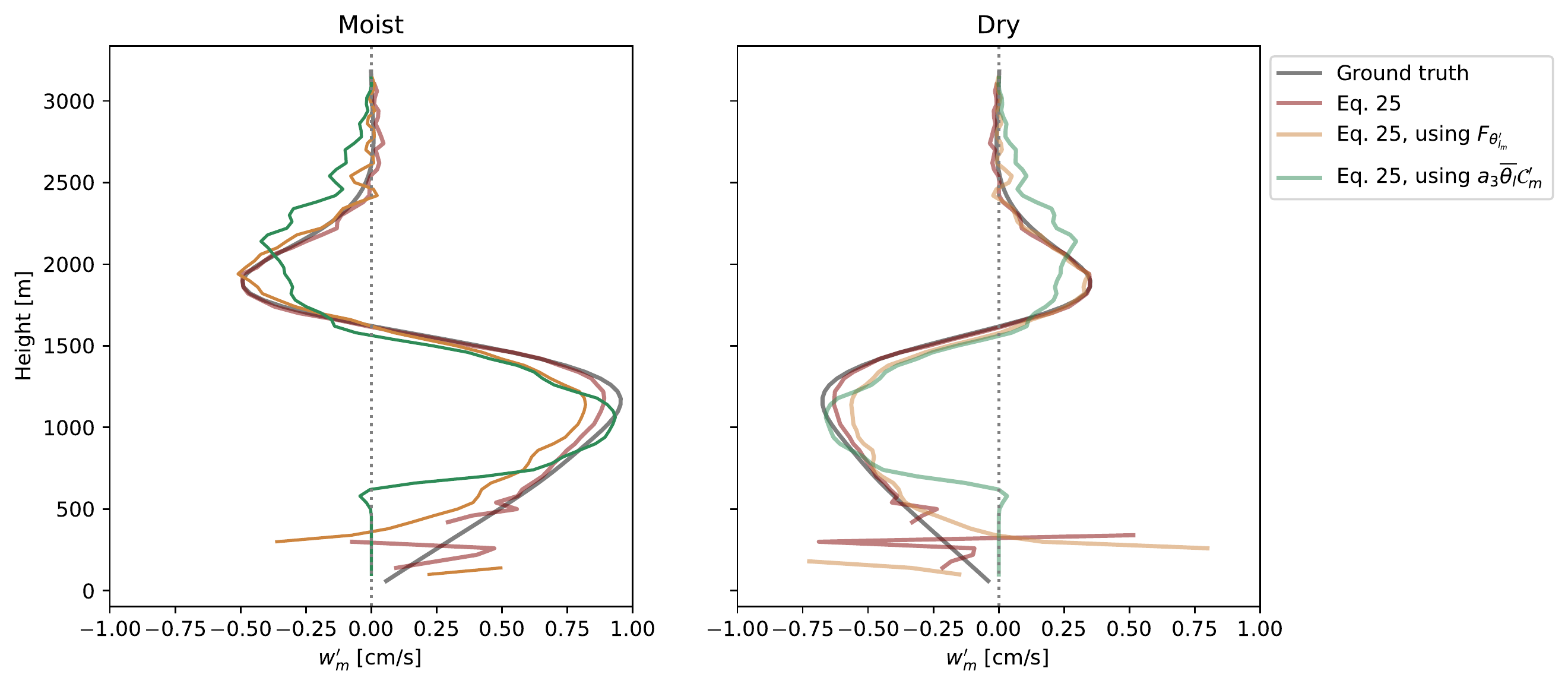}
    \caption{$w_m'$ derived directly from the LES model (black line) and modelled by eq.~\ref{e:wwtg} (maroon line), averaged over moist and dry regions and over 10-16 hr. Note also the comparatively small error made if $F_{{\theta_l'}_m}$ or $a_3\overline{\theta_l}\mathcal{C}_m'$ is used instead of $F_{{\theta_{lv}'}_m}$ in eq.~\ref{e:wwtg} (yellow and sea green lines).}
    \label{f:wpf_wtg}
\end{figure}

Fig.~\ref{f:wpf_wtg} a) and b) confirm the accuracy of this approximation, except in the sub-cloud layer where the gradients we divide through are too small for our simulations to resolve.

\subsubsection{The role of condensation}
What governs the vertical convergence of $F_{{\theta_{lv}'}_m}$? Fig.~\ref{f:thlvpf_budget} shows that it is positive in the moist region's cloud layer, and negative in the corresponding inversion layer. To explain this, we will for a final time return to our discussion from section~\ref{s:ana}. 

\begin{figure}[t]
    \centering
    \includegraphics[width=0.9\textwidth]{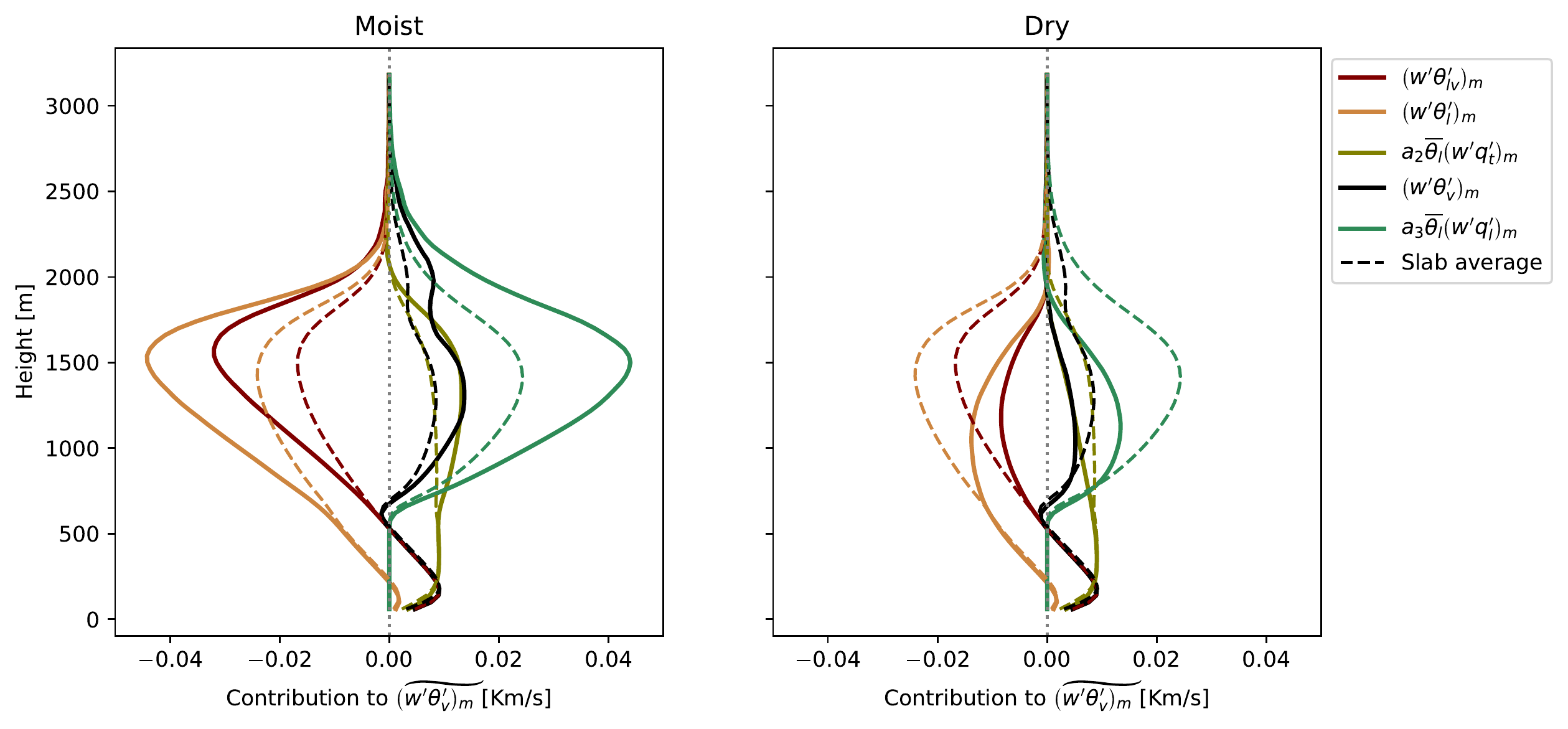}
    \caption{Grid-resolved, vertical mesoscale-filtered fluxes $\left(w'\theta_{v}'\right)_m$ and $\left(w'\theta_{lv}'\right)_m$ and their contributions from $\left(w'\theta_{l}'\right)_m$, $\left(w'q_{t}'\right)_m$ and $\left(w'q_{l}'\right)_m$ as defined in eq.~\ref{e:wthv}, averaged over moist and dry mesoscale regions (unbroken lines) and over the entire horizontal slab (dashed lines, copied from fig.~\ref{f:wthlvpav}), over 10-16 hr. The region-averaged flux anomaly $\widetilde{F_{\chi_m'}}$ (eq.~\ref{e:fchi}) is the difference between the unbroken and dashed lines for each variable.}
    \label{f:wthvp}
\end{figure}

Fig.~\ref{f:wthvp} plots the mesoscale-filtered and slab-averaged contributions to $w'\theta_{lv}'$, obtained from mesoscale filtering eq.~\ref{e:wthv}, and compares it to the slab-averaged fluxes previously presented in fig.~\ref{f:wthlvpav}. The vertical structure of the fluxes in the cloud layer is qualitatively similar when averaging them over either the entire slab, the moist mesoscale regions, or the dry mesoscale regions; they only differ in their magnitude. Therefore, just like we found $\overline{w'\theta_{lv}'}$ to be supported by $\overline{w'q_l'}$, we recognise that $\left(w'\theta_{lv}\right)_m$ is supported by $\left(w'q_l'\right)_m$, when averaged over moist and dry regions. This is possible in spite of $\theta_{lv_m}'\approx\theta_{v_m}'$ throughout the boundary layer (fig.~\ref{f:thvp}), because eq.~\ref{e:qlpf} shows that mesoscale condensation anomalies do not accumulate as $q_{l}'$, but instead give rise to an anomalous divergence of liquid water flux in the moist cloud layer. The contribution from $q_{l_m}'$ to $\theta_{lv_m}'$ (eq.~\ref{e:buoyf}) is small; the contribution from $\left(w'q_l'\right)_m$ to $\left(w'\theta_{lv}'\right)_m$ (filtered eq.~\ref{e:wthv}) is not.

In moist regions, fig.~\ref{f:wthvp} shows that $F_{{q_l'}_m} >0$. Using eq.~\ref{e:qlpf}, we recognise this to result directly from $\mathcal{C}_m'>0$, which consequently explains most of the moist cloud layer's anomalous heating, i.e. why $-\partial F_{{\theta_{lv}'}_m}/\partial z > 0$ in the left panel of fig.~\ref{f:thlvpf_budget}. Eq.~\ref{e:wwtg} then shows that this heating is immediately compensated by mesoscale ascent along the mean stratification, as mandated by the WTG model. As a result of the slab mean-exceeding liquid water transport from the layer below, the moist inversion layer must also evaporate more liquid water than the slab average, i.e. $\mathcal{C}_m'<0$. Consequently, the moist inversion layer experiences an anomalous convergence of $\left(w'q_l'\right)_m$ and an associated anomalous evaporative cooling, i.e. $-\partial F_{{\theta_{lv}'}_m}/\partial z < 0$ in fig.~\ref{f:thlvpf_budget}. Just like in the cloud layer below, this cooling is quickly balanced by a negative $w_m'$.

Hence, we have arrived at the heart of the mechanism: Horizontal, mesoscale anomalies in the same vertical structure of the net condensation $\mathcal{C}$ that governs the slab-mean layer's evolution have, under WTG, as a consequence that they develop mesoscale vertical motion of a few centimetres per second. To demonstrate this explicitly, $w_m'$ remains accurately predicted even when substituting $a_3\overline{\theta_l}\mathcal{C}_m'$ for $F_{{\theta_{lv}'}_m}$ in eq.~\ref{e:wwtg} (green lines in fig.~\ref{f:wpf_wtg}). 

Note that if we proceed along similar lines as above using \citet{betts1973non}'s original view of the slab-averaged problem, i.e. using mesoscale anomalies in $w'\theta_l'$ instead of $w'\theta_{lv}'$ to explain $w_m'$, the analysis remains largely unchanged, because, following the discussion in section~\ref{s:ana}, $w'\theta_l'$ too is well-known to be chiefly governed by $\mathcal{C}$. Since this view ignores virtual effects on the evolution of $\theta_{v_m}'$ (see eq.~\ref{e:buoyf}), it is slightly less accurate, but remains adequate for predicting $w_m'$ (yellow lines in fig.~\ref{f:wpf_wtg}). We emphasise that also in this view, $w_m'$ remains rooted in latent heating.

\subsubsection{\citetalias{bretherton2017understanding}'s model}

\begin{figure}[t]
    \centering
    \includegraphics[width=0.9\textwidth]{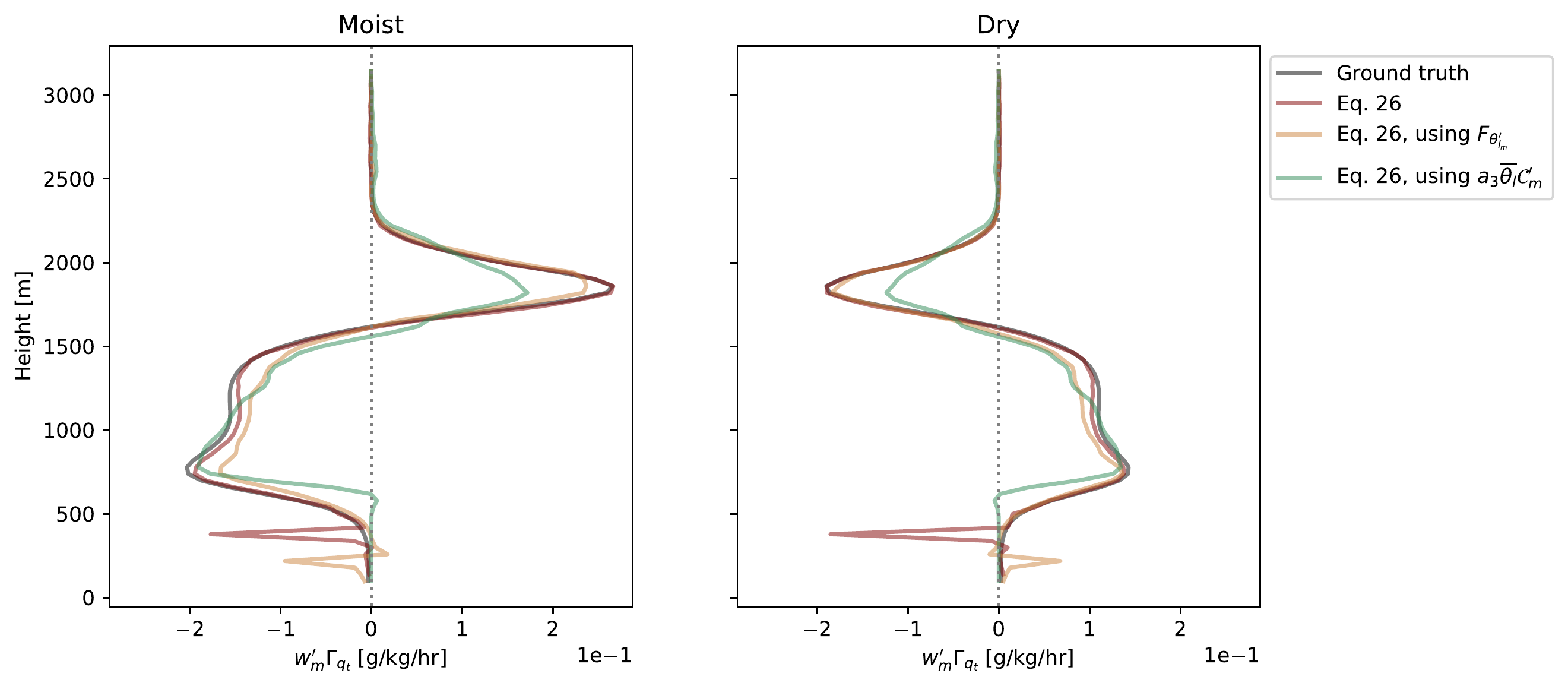}
    \caption{$w_m'\Gamma_{q_t}$ derived directly from the LES model (ground truth) following eq.~\ref{e:qtpfprodwtg} (maroon line), averaged over moist and dry regions and over 10-16 hr. Also shown are results if $F_{{\theta_l'}_m}$ or $a_3\overline{\theta_l}\mathcal{C}_m'$ is used instead of $F_{{\theta_{lv}'}_m}$ eq.~\ref{e:qtpfprodwtg} (yellow and sea green lines).}
    \label{f:qtpf_wtg}
\end{figure}

To complete the argument, one may multiply eq.~\ref{e:wwtg} with the negative, mean moisture gradient to finally arrive at a model for the onset of $q_{t_m}'$, formulated in terms of anomalous heat fluxes and the ratio of mean flow gradients

\begin{equation}
    \frac{\partial q_{t_m}'}{\partial t} \sim -w_m'\Gamma_{q_t} \approx \frac{1}{\rho_0}\frac{\partial}{\partial z}\left( \rho_0 F_{{\theta_{lv}'}_m} \right) \frac{\Gamma_{q_t}}{\Gamma_{\theta_{lv}}},
    \label{e:qtpfprodwtg}
\end{equation}

\noindent which fig.~\ref{f:qtpf_wtg} show is also accurate, and mostly captured even when replacing $F_{{\theta_{lv}'}_m}$ by $a_3\overline{a_3}\mathcal{C}_m'$ or $F_{{\theta_{l}'}_m}$. Eq.~\ref{e:qtpfprodwtg} is a succinct summary of the model presented by \citetalias{bretherton2017understanding}, combining the significant terms in their eq.20 and eq.32.

\section{Bulk model for the instability}
\label{s:add}

As \citetalias{bretherton2017understanding} note, eq.~\ref{e:qtpfiprodwtgmod} can in the language of tropical meteorology be understood as negative gross moist stability of moisture fluctuations \citep{neelin1987modeling,raymond2009mechanics}. Because of the absence of horizontal heterogeneity in radiation and precipitation in our simulations, we can here simplify the instability a little further than studies of deep convection typically do. In particular, we will close a simple, linear bulk instability model for the development of the moisture fluctuations, and examine the conditions of this model in some detail.


\subsection{Linear instability model}
\label{ss:fluxanom}

To close a positive feedback loop driving the development of $q_{t_m}'$, such fluctuations must lead directly to $F_{{\theta_{lv}'}_m}$, because vertical gradients in $F_{{\theta_{lv}'}_m}$ lead to mesoscale moistening anomalies following eq.~\ref{e:qtpfprodwtg}. Since $\mathcal{C}_m'$ supports $F_{{\theta_{lv}'}_m}$, an intuitive basis for this closure is to assume that a moister cloud layer is all that is needed for cumuli growing into it to condense more water. Fig.~\ref{f:concept} confirms that cloud-layer $q_{t_m}'$ and $\mathcal{C}_m'$ are well collocated. \citetalias{bretherton2017understanding} sketch a similar picture. However, a mathematical, theoretically founded description of the closure is still missing. Therefore, we suggest one here.

Our model will describe the time-evolution of bulk mesoscale moisture fluctuations, $\langle q_{t_m}'\rangle$. 
Vertical, partial integration of eq.~\ref{e:qtpfprodwtg} using eq.~\ref{e:vint} gives:

\begin{equation}
    \frac{\partial \langle q_{t_m}'\rangle}{\partial t} \approx 
    -\left. \frac{\rho_0}{\int_0^{z_\infty}\rho_0 dz} F_{{\theta_{lv}'}_m} \frac{\Gamma_{q_t}}{\Gamma_{\theta_{lv}}}\right|^{z_{\infty}}_0 - \left\langle F_{{\theta_{lv}'}_m} \frac{\partial}{\partial z}\left(\frac{\Gamma_{q_t}}{\Gamma_{\theta_{lv}}}\right) \right\rangle.
    \label{e:qtpfi1}
\end{equation}

$F_{{\theta_{lv}'}_m} = 0$ above the cloud layer and at the surface, courtesy of our lower boundary condition, setting the first term to zero. If we additionally assume that $\partial/\partial z\left(\Gamma_{q_t}/\Gamma_{\theta_{lv}}\right)$ is approximately constant with height, we may move it outside the integral:

\begin{equation}
    \frac{\partial \langle q_{t_m}'\rangle}{\partial t} \approx 
    - \frac{\partial}{\partial z}\left(\frac{\Gamma_{q_t}}{\Gamma_{\theta_{lv}}}\right) 
    \left\langle F_{{\theta_{lv}'}_m}\right\rangle.
    \label{e:qtpfi2}
\end{equation}

This assumption is not entirely accurate, but sufficiently reasonable through the cloud layer (not shown) that it is worth making in the interest of showing that bulk moistening of moist, mesoscale regions is governed by the integrated heat flux anomaly. Any parameterisation that relates such heat fluxes to $q_{t_m}'$ suffices to close the model, and many such parameterisations can be imagined. In the spirit of \citet{betts1975parametric}, we will merely use a simplified mass-flux approximation:

\begin{subequations}
    \begin{align}
    F_{{\theta_{lv}'}_m} & \approx -a_3\overline{\theta_{l}}F_{q_{l_m}'} \label{e:mod1}\\
    & \approx -C_1a_3\overline{\theta_{l}}w^*q_{l_m}' \label{e:mod2}\\
    & \approx -C\overline{\theta_{l}}w^*q_{t_m}'\label{e:mod3}.
    \end{align}
    \label{e:mod}
\end{subequations}

To write eq.~\ref{e:mod2}, we take $w^*$ to be a single vertical velocity averaged over all cloudy cells in both space and time, which ignores entrainment and detrainment effects, and which is less accurate than using cloud core variables \citep{siebesma1995evaluation}. However, it allows us to relate mesoscale fluctuations in in-cloud liquid water directly to $q_{l_m}'$ only. Since these assumptions yield errors only in the flux anomaly's magnitude, but not in the shape of its vertical profile \citep{siebesma1995evaluation}, we correct them with a second constant $C_1$. To write eq.~\ref{e:mod3}, we also assume $q_{l_m}'\propto q_{t_m}'$, and that equality can be restored by a single model constant $C$ which subsumes $C_1$ and $a_3$. This amounts to assuming that mesoscale fluctuations in the saturation specific humidity are small. 

\begin{figure}[t]
    \centering
    \includegraphics[width=0.8\linewidth]{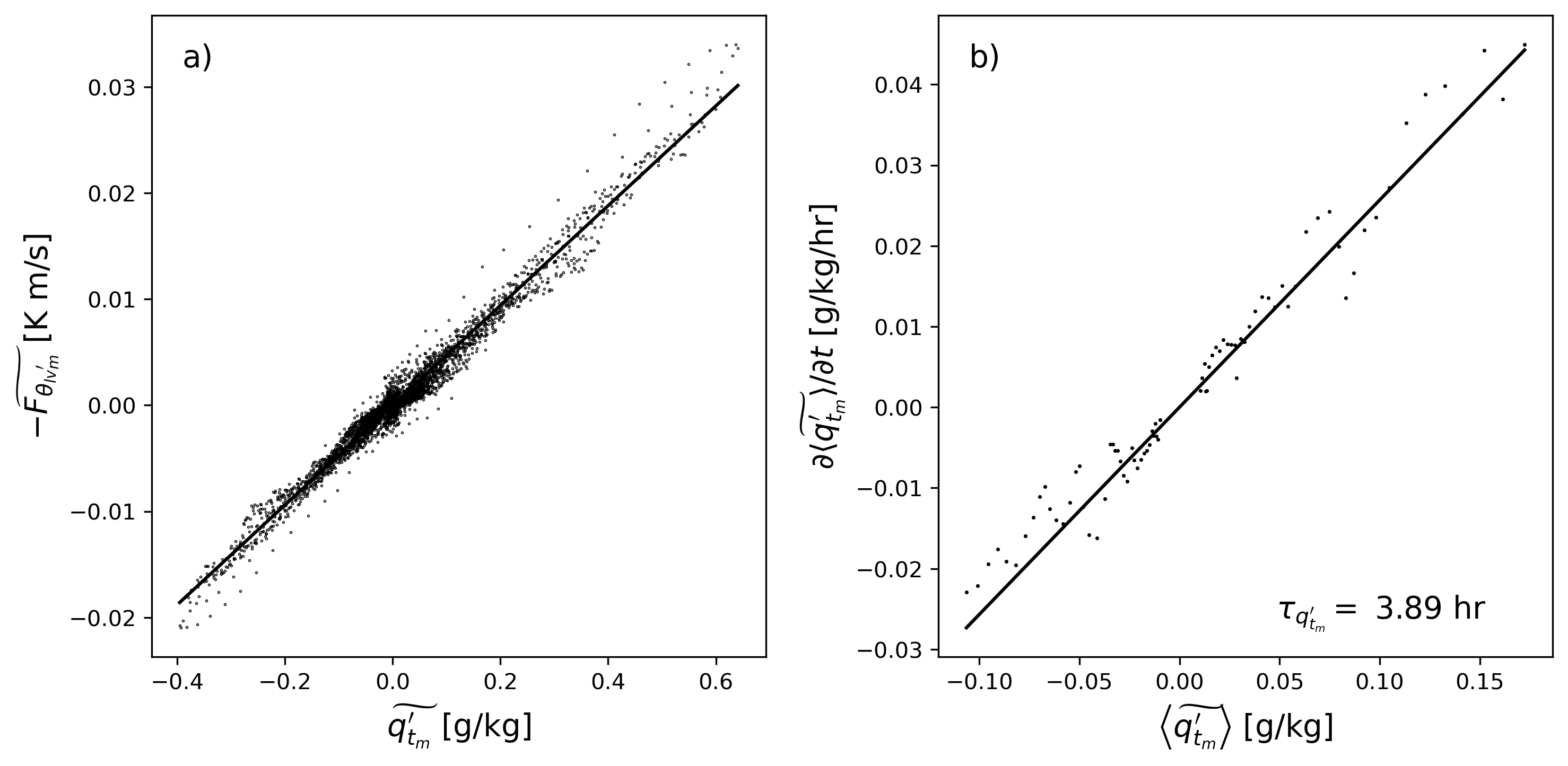}
    \caption{Scatterplots of moist and dry region averaged a) $q_{t_m}'$ against $-F_{{\theta_{lv}'}_m}$ and b) $\langle q_{t_m}'\rangle$ against $\partial\langle q_{t_m}'\rangle/\partial t$, for all fields between 6-16 hours (dots). Lines follow eq.~\ref{e:mod}, with constants $w^*=0.52$ m/s and $C=0.3$ (panel a), and eq.~\ref{e:qtpfiprodwtgmod}, with $\partial/\partial z\left(\Gamma_{q_t}/\Gamma_{\theta_{lv}}\right)=1.5\cdot10^{-3}$ g/kg/K/m (panel b). The constants are derived from the LES as described in the text. The time scale implied by the line in panel b (eq.~\ref{e:timescale}) is annotated.}
    \label{f:qtpf_wthlvpf}
\end{figure}

In spite of all these assumptions, we consider eq.~\ref{e:mod3} with $C=0.3$ adequate for the present discussion, c.f. fig.~\ref{f:qtpf_wthlvpf} a). Inserting this relation in eq.~\ref{e:qtpfi2} allows framing the growth of column-averaged mesoscale moisture fluctuations as a linear instability problem, whose time scale is $\tau_{q_{t_m}'}$:

\begin{subequations}
    \begin{equation}
        \frac{\partial \langle q_{t_m}'\rangle}{\partial t} \approx  \frac{\left\langle q_{t_m'} \right\rangle}{\tau_{q_{t_m}'}},
    \label{e:qtpfiprodwtgmod}
    \end{equation}
    \begin{equation}
        \tau_{q_{t_m}'} = \frac{1}{ C \overline{\theta_l} w^* \frac{\partial}{\partial z}\left(\frac{\Gamma_{q_t}}{\Gamma_{\theta_{lv}}}\right) }.
        \label{e:timescale}
    \end{equation}
    \label{e:instabmodel}
\end{subequations}

Eq.~\ref{e:instabmodel} remains rather accurate (fig.~\ref{f:qtpf_wthlvpf} b), diagnosing a time scale for the instability of almost 4 hours in our simulation. The model is also plotted in fig.~\ref{f:qtpf_budget_int}. 

While illustrative, it is prudent to ask if eq.~\ref{e:mod}, upon which this time scale estimate rests, is reliable. Since it depends heavily on $w^*$, which is not well-constrained by any argument we have made, but is energetically supported by in-cloud turbulence at the very smallest scales our numerical model resolves, this is in fact quite questionable. \citetalias{bretherton2017understanding}, who estimate the time scale without reference to a model for it, obtain a significantly larger number (15 hours) than we do, suggesting that the mechanism may exhibit a strong numerical dependence, as is observed for models of self-aggregating deep convection \citep[e.g.][]{muller2012detailed,wing2020clouds}. We devote a separate manuscript to this issue.

\subsection{Condition for instability}
\label{ss:gradrat}

If the assumptions made in deriving eq.~\ref{e:instabmodel} are generally valid, its only condition for unabated growth of $q_{t_m}'$ is that $\partial /\partial_z \left(\Gamma_{q_t}/\Gamma_{\theta_{lv}}\right) > 0$. \citetalias{bretherton2017understanding} arrive at the same condition, but formulate it as a demand that a convex relation must exist between $\overline{q_t}(z)$ and $\overline{\theta_{lv}}(z)$ (or another pair of conserved thermodynamic variables). This follows from rewriting

\begin{equation}
    \frac{\partial}{\partial z}\left(\frac{\Gamma_{q_t}}{\Gamma_{\theta_{lv}}}\right) = \Gamma_{\theta_{lv}}\frac{\partial^2 \overline{q_t}}{\partial \overline{\theta_{lv}}^2}
    \label{e:instabreq}
\end{equation}

\noindent with the quotient rule of calculus. Hence, a stable stratification ($\Gamma_{\theta_{lv}}>0$) and $\partial^2 \overline{q_t}/\partial \overline{\theta_{lv}}^2>0$ are needed to translate $F_{{\theta_{lv}'}_m}$ into a positive, column-averaged gradient production of $q_{t_m}'$. This requirement arises because the divergence of $F_{{\theta_{lv}'}_m}$ itself integrates to zero. Were there a linear, mixing-line relation between $\overline{\theta_{lv}}$ and $\overline{q_t}$, the gradient production of $q_{t_m}'$ would integrate to zero also. In contrast, a concave relation between $\overline{\theta_{lv}}$ and $\overline{q_t}$ would result in a self-stabilising feedback, where heating fluctuations would lead to column-drying. 

Our simulations' initial profiles of $\overline{q_t}$ and $\overline{\theta_{lv}}$ are straight lines, i.e. initially $\partial^2 \overline{q_t}/\partial \overline{\theta_{lv}}^2=0$. Any convexity in this relation must consequently develop in time. \citetalias{bretherton2017understanding} suggest that the convexity may be brought about by large-scale cooling and advection. However, by deriving an evolution equation for $\partial^2 \overline{q_t}/\partial \overline{\theta_{lv}}^2$, one can show that in LES simulations where the thermodynamic state, subsidence and large-scale forcing profiles are initially at most linear functions of height, these processes cannot induce $\partial^2 \overline{q_t}/\partial \overline{\theta_{lv}}^2 > 0$ (we present the full derivation in appendix B). Instead, it must develop from slab-averaged fluxes of $\theta_{lv}$ and $q_t$, whose vertical structure must satisfy

\begin{subequations}
    \begin{equation}
        \frac{1}{\rho_0}\frac{\partial^3}{\partial z^3}\left(\rho_0\overline{w'q_t'}\right) < 0
        \label{e:divcurvq}
    \end{equation}
    \begin{equation}
        \frac{1}{\rho_0}\frac{\partial^3}{\partial z^3}\left(\rho_0\overline{w'\theta_{lv}'}\right) < 0
        \label{e:divcurvt}
    \end{equation}
    \label{e:divcurv}
\end{subequations}

\noindent when $\Gamma_{\theta_{lv}}>0$ and $\Gamma_{q_t}<0$. 

Once $\overline{\theta_{lv}}$ and $\overline{q_t}$ are no longer linear functions of height, these constraints can be relaxed to include effects from linear profiles of subsidence and large-scale forcing, and curved profiles of the heat and moisture fluxes. However, since BOMEX features large-scale forcings in the cloud layer that are \textit{constant} in height, these processes are prohibited from having \textit{any} effect on the development of convexity in our cloud layer's mean state. Hence, in contrast to what \citetalias{bretherton2017understanding} suggest, we find that the inception of $\partial^2 \overline{q_t}/\partial \overline{\theta_{lv}}^2>0$ is not imposed by large-scale forcing, but is an internal feature of our simulated, slab-averaged cumulus convection's vertical structure.

\begin{figure}[t]
    \centering
    \includegraphics[width=0.6\linewidth]{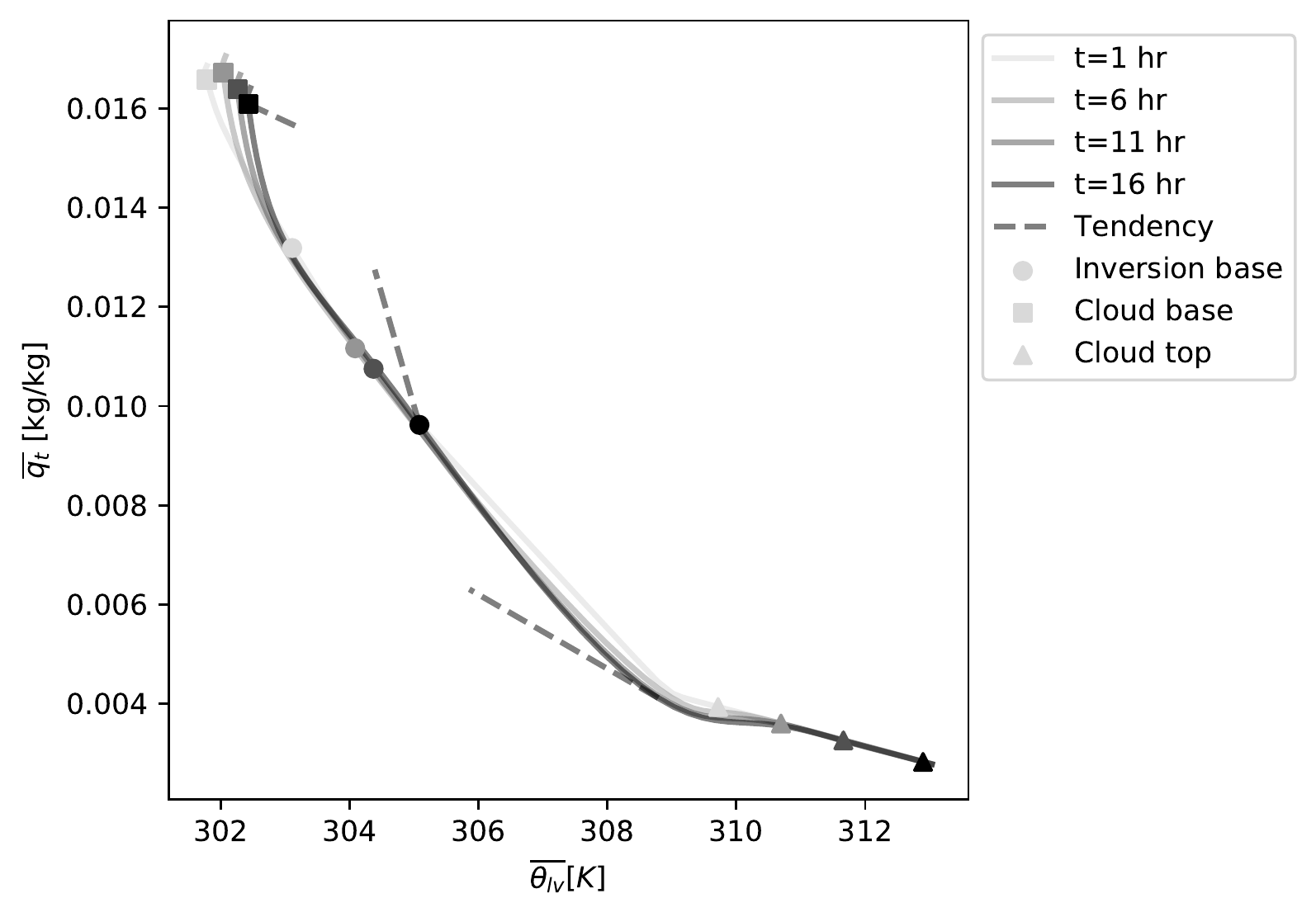}
    \caption{Development of the vertical relation between $\overline{\theta_{lv}}$ and $\overline{q_t}$ over four time samples, superimposed by their tendency at the last point in time, at several altitudes. Circles indicate the inversion base, squares the cloud base, and triangles cloup top, as they vary with time.}
    \label{f:conserved}
\end{figure}

\begin{figure}[t]
    \centering
    \includegraphics[width=0.9\linewidth]{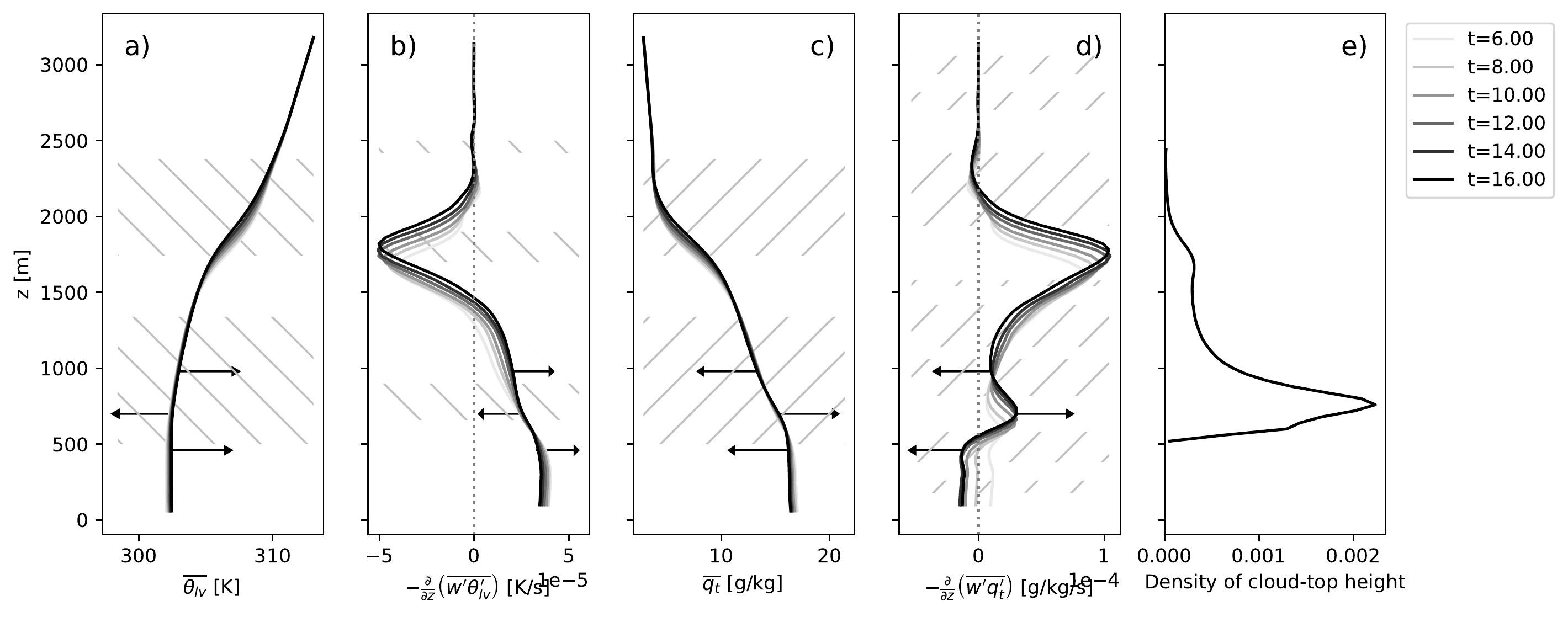}
    \caption{Time-evolution of vertical profiles of $\overline{\theta_{lv}}$ (a) and $\overline{q_t}$ (b), $\frac{\partial}{\partial z}\left(\overline{w'\theta_{lv}'}\right)$ (c) and $\frac{\partial}{\partial z}\left(\overline{w'q_t'}\right)$ (d), and vertical distribution of cloud-top height throughout the simulation (e). Hatches indicate areas where $\partial^2 \overline{q_t} / \partial \overline{\theta_{lv}} ^2>0$ (a and b), $\partial^3 \overline{w'\theta_{lv}'} / \partial z^3<0$ (c) and $\partial^3 \overline{w'q_t'} / \partial z^3<0$ (d) at $t=16$ hr. Arrows denote the action of condensation and evaporation in very shallow clouds over the transition layer.}
    \label{f:convexity}
\end{figure}

In our simulations, the initially linear profiles of $\overline{q_t}$ and $\overline{\theta_{lv}}$ acquire a convex relation in the lower cloud layer and in the inversion (fig.~\ref{f:conserved}, hatched regions in fig.~\ref{f:convexity} a and b). This structure is generated by slab-mean heating and drying in the lower cloud layer, and moistening and cooling in the inversion layer (dashed lines in fig.~\ref{f:conserved}); these tendencies can in turn largely be attributed to slab-mean flux convergence of heat and moisture (fig.~\ref{f:convexity} c and d, see also figs.~\ref{f:slabav} a and b). The hatched regions in fig.~\ref{f:convexity} c and d show that their vertical structure satisfies eq.~\ref{e:divcurv} precisely where $\partial^2 \overline{q_t}/\partial \overline{\theta_{lv}}^2>0$, as expected.

A rather elegant explanation for this fact is suggested by \citet{albright2022}. In observations, the cloud-top height distribution features two modes surrounded by a rather wide spread: One due to very shallow cumuli between 500m-1000m and one due to deeper clouds, which reach the inversion layer between 1600m and 2000m. Our simulations exhibit the same bimodality (fig.~\ref{f:convexity} e)), with the modes coinciding precisely with the layers where $\partial^2\overline{q_t}/\partial{\overline{\theta_{lv}}}>0$.

The first mode spans the so-called transition layer, which we loosely interpret as the layer where $\overline{q_t}$ is no-longer well-mixed, while $\overline{\theta_{lv}}$ is. Observations both old \citep{augstein1973mass} and very recent \citep{albright2022subcloud} indicate that thick, curved transition layers are omnipresent in the trades. \citet{albright2022} suggest that such a structure may be brought about by the population of very shallow clouds that inhabit this layer (see arrows in fig.~\ref{f:convexity}): At cloud base, they warm and moisten in accordance with deeper clouds. Yet, they quickly evaporate (around 700m), yielding a cooling and moistening that is analogous to how deeper clouds cool and moisten the inversion, and that is visible in the flux convergence profiles. Once the very shallow clouds dissipate, the positive net condensation in the remaining, deeper clouds returns the heating and moistening the layer. This is what brings about the structure in the heat and moisture fluxes that ensure $\partial^2\overline{q_t}/\partial{\overline{\theta_{lv}}}>0$. Similarly, the thickness and curvature of the inversion layer is brought about by variability in how deep into that layer individual, deeper cumuli reach, and thus where they evaporate, cool and moisten the layer.

Demanding curvature of the flux convergence profiles takes us two orders beyond quasi-steady assumptions that sometimes feature in simple models of cumulus layers \citep{stevens2007growth}; these require the slab-mean flux divergence to be constant with height. Yet, above our simulation's well-mixed layer, gradients and curvature in the flux divergence are the norm, rather than the exception, especially near cloud base and the inversion. In fact, the resulting, subtle deviations from linearity in the vertical structure of $\overline{\theta_l}$ and $\overline{q_t}$ are a ubiquitous feature across the canon of LESs that exhibit length scale growth in their trade-wind cumulus layers, e.g. the original Rain in Cumulus over Ocean (RICO) ensemble \citep{vanzanten2011controls} and its derivatives \citep[e.g.][]{seifert2015large,anurose2020understanding}, the idealised framework developed by \citet{bellon2012using} and later used for studies of cloud organisation by \citet[e.g.][]{vogel2016role}, the simulations of length scale growth presented by \citet{narenpitak2021sugar} and even the simulations \citep{blossey2013marine} that \citetalias{bretherton2017understanding} develop their theory upon, in spite of all these cases including much more detailed physics than we do here. The condition eq.~\ref{e:divcurvq} is satisfied even for the moist static energy fluxes diagnosed over the undisturbed BOMEX period by \citet{nitta1974heat}; their satellite images also indicate that large cloud structures developed even during the undisturbed period so long associated only with small, stable cumuli. 

In all, we find rather convincing evidence that the conditions required to destabilise mesoscale moisture fluctuations are usually satisfied, and that it is not necessary to resort to explanations involving the larger-scale forcing on the system. In making this statement, we are essentially arguing for the more general validity of  \citetalias{bretherton2017understanding}'s results: Shallow cumulus convection is intrinsically unstable to length scale growth. 

\section{Discussion and outlook}
\label{s:dis}

Before summarising, let us review three consequences of this rather striking conclusion.

\subsection{Relevance of circulation-driven scale growth}

To what extent does condensation-driven scale growth matter in nature? To lay bare the essence of the mechanism, we have here made a number of simplifications that are probably too restrictive for us to speak authoritatively on this matter. In particular, our assumptions that i) the surface fluxes and large-scale advection of heat and moisture are fixed in space and time, ii) precipitation and interactive radiation do not play a role and iii) the mean environment is rather stationary, imply that we ignore several important processes that in nature will modulate the instability we discuss. Since the relative effects of such processes in patterning the trades is a topic of active research, we briefly discuss some anticipated consequences of these assumptions here. 

First, if $q_{t_m}'$ reach the surface, eq.~\ref{e:qtpfi1} shows that surface flux anomalies could potentially oppose further moistening. At constant sea surface temperatures and with mild precipitation, \citetalias{bretherton2017understanding} show such effects to be of second order importance, indicating that the well-mixed layer may remain quasi-stationary and tied to the local surface conditions at large spatial scales, as is often assumed in models of deep convective self-organisation \citep[e.g.][]{emanuel2014radiative}. This matters, since sea-surface temperature influences the vigour of the convection (measured here through $w^*$) and thus, through eq.~\ref{e:instabmodel}, can explain e.g. the more rapid production of mesoscale circulations and moist patches over a warmer ocean simulated by \citet{vogel2016role}. Extensions to also study sea-surface temperature heterogeneities, which amplify circulations \citep{park2006modification}, are easily imagined.

Heterogeneous radiation can support shallow circulations in detailed \citep{klinger2017effects} and conceptual \citep{naumann2019moist} simulations of cumulus-topped boundary layers. In particular, when the circulations are sufficiently strong to begin detraining significant amounts of inversion cloud atop the moist region's boundary layer, they reinforce the anomalous heating that here drives the circulations \citepalias{bretherton2017understanding,vogel2020influence}; these effects would accelerate the mechanism beyond the time-scale eq.~\ref{e:timescale}.

If the shallow cumulus layer deepens sufficiently for precipitation to form, we must further amend our estimates. Slab-averaged precipitation will on one hand reinforce $\Gamma_{q_t}<0$ in the cloud layer, but on the other reduce the inversion-layer evaporation upon which $F_{{\theta_{lv}}_m}$ relies \citep{albrecht1993effects}; the relative effects of these factors seem to enhance length scale growth in the simulations conducted by \citetalias{bretherton2017understanding}. However, precipitation that is sufficiently vigorous to produce cold pools will locally discourage the formation of convection upon which our simulated circulations rely, and may thus relieve gradient production driven by cloudy updrafts as the leading-order dynamics that govern the spatial distribution of moisture and clouds; such transitions seem to take place in mesoscale LES of RICO \citep{seifert2013large,seifert2015large,anurose2020understanding,thomas2021toward}. When cold pools dominate, their length scale replaces that of the clouds as the descriptor of the mesoscales \citep{matheou2021growth}, and the length scales of cloud-free areas emerge as the natural complement to those of the clouds themselves as measures of the resulting cloud organisation \citep{janssens2021cloud,schulz2021characterization}. However, a broad regime of conditions can be imagined where both circulations and cold pools act in concert to pattern the convection, e.g. under strong inversions and high surface fluxes, as for the patterns labelled Flowers and Fish in \citet{schulz2021characterization}, or through an imposed large-scale ascent which may vary from being upwards and later downwards \citep{narenpitak2021sugar}. Such situations deserve more study.

Finally, we note that recent observations \citep{george2021observed,george2021joanne} suggest that mesoscale circulations with a similar magnitude and vertical structure as we find in our simulations pervade the trades. Also, we note that any process which gives rise to mesoscale circulations will be amplified by the mechanism discussed here. Hence, while more research is needed to explicitly root these observations in the dynamics described here and by \citetalias{bretherton2017understanding}, the mechanism should be a central hypothesis in any study attempting to explain mesoscale variability in clouds and moisture.

\subsection{Connection to trade-inversion growth}

Application of the WTG approximation implies that the level of the trade inversion is rather constant over our domain, in spite of large moisture fluctuations accumulating over the layer. The level of this trade-inversion is governed by $w'\theta_{lv}'$ \citep[section~\ref{s:ana};][]{stevens2007growth}, fluctuations in which also govern scale growth. Combining these observations gives two insights.

First, it highlights a very practical way in which mesoscale cloud fluctuations affect the slab-averaged layer: Since $\overline{w'\theta_{lv}'}$ skews towards the profiles set by the deeper clouds in moist regions (see fig.~\ref{f:wthvp}), the scale growth raises the inversion. This explains why, in contrast to \citet{siebesma1995evaluation}'s small-domain simulations, our domain-averaged thermodynamics are not steady (fig.~\ref{f:slabav}). Unless balanced by an increased subsidence with height or precipitation, scale growth and deepening of the inversion go hand in hand.

The upshot is that the self-aggregation of moisture discussed here may influence transitions to deep convection. In fact, after around 20 hours, the moist patches in our simulation develop deep, organised clouds, aided by the lack of subsidence above 2 km in our simulation setup. This feedback is similar to that described by \citet{jeevanjee2013convective}; the fundamental instability to scale-growth of shallow convection may thus also initiate the self-organisation of deep convection.

\subsection{Connection to cloud feedback estimates}

How does the scale growth mechanism affect cloud fraction, which to first order govern the trades' contribution to the equilibrium climate sensitivity, and which remains poorly constrained in general circulation models \citep{zelinka2020causes}? Fig.~\ref{f:twp_cld_evo} shows the cloud fraction is remarkably robust over our 16 hours of simulation, as cloudiness increases in moist regions compensate reductions in dry regions. The small, observed reduction can be attributed to the developing circulation's tendency to contract the moist regions at the expense of the dry regions (fig.~\ref{f:qtpf_budget_int}), an observation which is consistent with e.g. \citet{vogel2016role}.

The weak dependence of the cloud fraction on the cloud length scale is consistent with observations \citep{janssens2021cloud}, which indicate a rather weak relation between measures of a cloud field's characteristic length or aggregtion, and cloud fraction. Even if the mechanism would strengthen above warmer sea surfaces or in more weakly subsiding environments, it would thus likely support the emerging picture that trade-wind cloudiness is rather insensitive to changes in the overall climate \citep{myers2021observational,cesana2021observational}.

However, two notes on this statement motivate further research. First, the cloud fraction will be sensitive to the developing inversion-layer outflows' ability to sustain extensive sheets of inversion cloud, which it does not do in our simulations, but which is observed in other studies \citep{bretherton2017understanding,vogel2020influence,narenpitak2021sugar,bony2020sugar}. Many situations can be imagined to feature higher inversion cloud fractions than BOMEX, whose inversion is rather dry and warm. More systematic study of the mechanism over different environmental conditions using more realistic physics than we do here is warranted.

This is particularly pertinent because approaches such as those taken by \citet{myers2021observational,cesana2021observational} essentially assume large-scale cloud-controlling variables set the cloud fraction. Recent observations lend credence to this approach, suggesting that it is (presumably externally induced) variability in vertical velocity at larger-scales which raises the subcloud layer and thus gives rise to stronger cloud-base mass fluxes and cloud fractions \citep{bony2019measuring,vogel2020estimating,george2021observed}.

Our results, however, suggest the opposite view: Here, spatial variability in the convective mass flux, filtered and averaged over mesoscale moist and dry regions, give rise to variability in larger-scale vertical velocity. The role of the resulting circulation is to set the right cloud-layer thermodynamic environment for clouds to preferentially form in, which results in the stronger mass fluxes in regions of positive, large-scale vertical velocity. We also do not observe the subcloud layer height to differ appreciably between moist and dry spatial regions (see fig.~\ref{f:concept}). If the view suggested by our simulations turns out to matter in nature, questions arise regarding the validity of approaches such as those taken by \citet{myers2021observational,cesana2021observational}, because they ignore that shallow convective clouds may simply control their own cloud-controlling variables. Reconciling the views put forward on the basis of recent observations with ours is thus a recommendation with substantial ramifications \citep{bony2015clouds}. Fortunately, the data from the recent EUREC\textsuperscript{4}A field campaign \citep{bony2017eurec4a,stevens2021eurec} may be sufficiently detailed to begin answering such questions, boding well of our understanding of the significance of self-organising shallow cloud patterns to climate.

\section{Summary and concluding remarks}
\label{s:sum}

Building on \citet{bretherton2017understanding}, we have formulated an idealised model for a linear instability that leads to uninhibited length scale growth of moisture fluctuations in layers of non-precipitating trade-wind cumulus (eq.~\ref{e:instabmodel}). Using only well-established theory and a classical large-eddy simulation setup (BOMEX) with no heterogeneous surface forcing, radiation or precipitation, the model explains how small spatial differences in the amount of latent heating in shallow cumulus clouds produce a mesoscale circulation under the assumption of weak, horizontal mesoscale temperature gradients. The circulation converges moisture into regions that consequently support more cumulus clouds, diabatic heating and a stronger circulation; these regions grow exponentially in intensity and scale (fig.~\ref{f:qtpf_budget_int}) until they are modulated by an outer length scale (here the finite size of our LES domains) or translate the problem to a regime of different leading-order dynamics, e.g. driven by precipitation or radiation, which we do not simulate.

Contrary to what \citetalias{bretherton2017understanding} suggest, in our idealised LES simulations the imposed environment of a scale larger than the domain does not directly influence the convective instability, as long as it supports a mean cumulus layer; the instability is free to develop on top of the mean state as a function only of turbulent fluxes of heat and moisture (eq.~\ref{e:divcurv}), because they naturally give rise to curvatures in the mean state. 

In all, we conclude that shallow convection is intrinsically unstable to scale growth, a result which is implied even by the results reported by \citet{nitta1974heat} for the ``undisturbed'' BOMEX period, upon which many ideas that assume horizontal homogeneity in trade-wind cloudiness rely. Our results should aid to disassemble the last vestiges of such ideas.

As a final remark, we note how striking it is that we have only required well-established, classical theory for our discussion. As noted at the outset, the structure of the mean trades was elucidated sixty years ago. WTG's utility has been known for forty years in the community of tropical meteorology \citep{held1985large}. Even the interpretation of the instability we have discussed as negative values in moist gross stability \citepalias{bretherton2017understanding}, relates to classical, influential concepts from tropical meteorology \citep{neelin1987modeling}. This motivates us to conclude simply by asking: What else might we learn from the insights of the giants of tropical meteorology when exploring the rather uncharted territory of the mesoscale trades?

\clearpage
\section*{Acknowledgments}

MJ warmly acknowledges illuminating conversations with Anna Lea Albright on the role of very shallow cumulus in shaping the transition layer. CvH acknowledges funding from the Dutch Research Council (NWO) (grant: VI.Vidi.192.068). A. Pier Siebesma acknowledges funding by the European Union’s Horizon 2020 research and innovation program under grant agreement no. 820829 (CONSTRAIN project). FG acknowledges support from The Branco Weiss Fellowship - Society in Science, administered by ETH Z\"urich, and from an NWO Veni grant. Finally, we thank the Dutch Reserach Council NWO for use of its computer facilities (project 2021/ENW/01081379).

\section*{Open science}

To ensure the reproducibility of the results, we make public the version of DALES \\(\url{https://doi.org/10.5281/zenodo.6545655}), the numerical settings \\ (\url{https://doi.org/10.6084/m9.figshare.19762219.v1}), and routines used to generate the plots presented herein (\url{https://doi.org/10.5281/zenodo.6545917}). Living repositories for DALES and the postprocessing scripts are available at \\ \url{https://github.com/dalesteam/dales} and \url{https://github.com/martinjanssens/ppagg}.

\appendix

\section{Budgets of scalars}
\label{app:a}

\subsection{Derivation of eq.~\ref{e:localbudgrew}}

Eq.~\ref{e:localbudgrew} in the main text may be derived from eq.~\ref{e:localbudg} by making use of the decomposition into slab-averaged $\overline{\cdot}$ and fluctuating $\cdot'$ quantities (eq.~\ref{e:fluc}), yielding the following expansions for the horizontal and vertical advection terms:

\begin{subequations}
    \begin{align}
        \frac{\partial}{\partial x_{jh}}\left(u_{jh}\chi\right) = & \frac{\partial}{\partial x_{jh}}\left(\overline{u_{jh}}\chi + 
        u_{jh}'\overline{\chi} +
        u_{jh}'\chi'
        \right) \\
        = & \chi \frac{\partial \overline{u_{jh}}}{\partial x_{jh}} + 
        \overline{\chi} \frac{\partial u_{jh}'}{\partial x_{jh}} + 
        \overline{u_{jh}} \frac{\partial \chi}{\partial x_{jh}} + 
        u_{jh}' \frac{\partial \overline{\chi}}{\partial x_{jh}}
        \label{e:horadv}
    \end{align}
    \begin{align}
        \frac{1}{\rho_0}\frac{\partial}{\partial z}\left(\rho_0 w\chi \right) = & \frac{1}{\rho_0}\frac{\partial}{\partial z}\left(\rho_0 \left(
        \overline{w}\chi + 
        w'\overline{\chi} +
        w'\chi' \right)\right) \\
        = &  \chi\left(
            \frac{\partial\overline{w}}{\partial z} + 
            \frac{1}{\rho_0}\frac{\partial \rho_0}{\partial z}\overline{w} \right) +
        \overline{\chi}\left(
            \frac{\partial w'}{\partial z} +
            \frac{1}{\rho_0}\frac{\partial \rho_0}{\partial z}w'\right) +
        \overline{w}\frac{\partial \chi}{\partial z} + 
        w'\frac{\partial \overline{\chi}}{\partial z} +
        \frac{1}{\rho_0}\frac{\partial}{\partial z}\left(\rho_0 w'\chi' \right)
        \label{e:veradv}
    \end{align}
\end{subequations}
In the anelastic approximation, conservation of mass demands:

\begin{equation}
    \frac{\partial u_{jh}}{\partial x_{jh}} + \frac{\partial w}{\partial z} + \frac{1}{\rho_0}\frac{\partial \rho_0}{\partial z}w = 0
    \label{e:anelmass}
\end{equation}

\noindent with the last term required to conserve the reference mass \citep{lilly1996comparison}; it holds for both fluctuations and in the slab-average. Therefore, when adding the expansions eq.~\ref{e:horadv} and eq.~\ref{e:veradv}, the respective sum of the first terms and second terms in these equations (those scaled by $\chi$ and $\overline{\chi}$) are zero, resulting in eq.~\ref{e:localbudgrew} of the main text. Note that because we solve our equations on a doubly periodic domain, eq.~\ref{e:anelmass} requires $\overline{w}=0$, its effects in eq.~\ref{e:localbudgrew} are thus prescribed by setting $\overline{w}$ and scaling it with the local vertical gradient of $\chi$.

\subsection{Derivation of eq.~\ref{e:chipf} for mesoscale scalar fluctuations}

The main text's eq.~\ref{e:chipf} can be derived from eq.~\ref{e:localbudgrew} and eq.~\ref{e:slabav} by subtracting the latter from the former, which gives:

\begin{equation}
    \frac{\partial \chi'}{\partial t} =
    -\overline{u_{jh}}\frac{\partial \chi'}{\partial x_j}
    -u_{jh}'\frac{\partial\overline{\chi}}{\partial x_{jh}} -\frac{\partial}{\partial x_{jh}}\left( u_{jh}'\chi'
    - \overline{u_{jh}'\chi'} \right)
    - \overline{w}\frac{\partial \chi'}{\partial z}
    - w'\frac{\partial \overline{\chi}}{\partial z}
    - \frac{1}{\rho_0}\frac{\partial}{\partial z}\left(\rho_0 \left(w'\chi' - \overline{w'\chi'}\right)\right)  
    \label{e:chip}
\end{equation}

To derive eq.~\ref{e:chipf} from eq.~\ref{e:chip}, we have made a number of assumptions. In our LES model, whose results we analyse, use of doubly periodic boundary conditions enforces the following conditions on the horizontal advection of fluctuations:

\begin{subequations}
    \begin{equation}
        \overline{u_{jh}}\frac{\partial \chi'}{\partial x_j} = \frac{\partial}{\partial x_{jh}}\left(\overline{u_{jh}}\chi'\right)
        \label{e:horadvmean}
    \end{equation}
    \begin{equation}
        u_{jh}'\frac{\partial \overline{\chi}}{\partial x_{jh}} = 0
        \label{e:horadvfluc}
    \end{equation}
    \begin{equation}
        \frac{\partial}{\partial x_{jh}}\left(\overline{u_{jh}'\chi'}\right) = 0
        \label{e:horadvflux}
    \end{equation}
\end{subequations}

Thus, unless their effects would be prescribed or parameterised, our model does not account for i) the advection of scalar fluctuations into the analysed domain with the mean wind (eq.~\ref{e:horadvmean}), ii) interactions between horizontal gradients in $\chi$ larger than the domain (eq.~\ref{e:horadvfluc}) or iii) horizontal eddy-fluxes into the domain (eq.~\ref{e:horadvflux}). For our idealised analysis of the onset of scale-growth from local processes, these assumptions seem reasonable, but probably become untenable for analyses of finite, real-world domains with open boundaries.

Furthermore, we have in our analysis neglected the explicit effects of unresolved-scales effects. These would enter the analysis through additional diffusion terms on the right-hand side of eq.~\ref{e:localbudg}. We do not present them in our equations, but we do compute them and include them in the appropriate flux divergence terms in the budgets presented in the text. At the mesoscales, which are far removed from their action on the smallest, resolved scales, their direct effects are small. Nevertheless, their influence in setting the fluxes which drive the model is non-trivial, as we will show in future work.

Making these assumptions and applying a mesoscale filter to the resulting equation results in eq.~\ref{e:chipf}, and ensures its consistency with our LES model.

\subsection{Moist and dry region averaging}
Eq.~\ref{e:chipf} is formulated in an Eularian manner. When averaging it over moist and dry regions, as done in the main text, we risk that our local budgets become dominated by mean-flow advection of the regions. Since we are more interested in the evolution of the regions themselves, we note that we can make use of Reynolds' transport theorem (here manipulated with the divergence theorem)

\begin{equation}
    \widetilde{\frac{\partial \chi_m'}{\partial t}} = \frac{\partial\widetilde{\chi_m'}}{\partial t} - \frac{\partial}{\partial x_{jh}}\widetilde{\left(\chi_m' u^b_{jh}\right)}
    \label{e:reynoldstrans}
\end{equation}

\noindent where $\widetilde{\cdot}$ represents the moist or dry region averaging operator and $u^b_j$ is the horizontal velocity of the region's boundary. The first term on the right-hand side of eq.~\ref{e:reynoldstrans} captures the evolution of $\chi_m'$ averaged over the moist and dry regions, in which we are primarily interested, and which is plotted in fig.~\ref{f:qtpf_budget} and ~\ref{f:thlvpf_budget}. The second term accounts for the advection and net expansion of the regions with $u^b_{jh}$. 
By decomposing $u^b_{j_h}$ into its contributions from mean flow advection ($\overline{u}_{jh}$) and net expansion ($u^e_{jh}$)

\begin{equation}
    u^b_{jh} = \overline{u}_{jh} + u^e_{jh},
\end{equation}


\noindent and inserting this into the second term on the right-hand side of eq.~\ref{e:reynoldstrans}, we recognise that we may cancel the mean-flow advection term that results with the mean-flow advection contribution to the second term on the right-hand side of eq.~\ref{e:chipf}, if it is expanded as follows:

\begin{equation}
    \frac{\partial}{\partial x_{jh}}\left(u_{jh}\chi'\right)_m = \frac{\partial}{\partial x_{jh}}\left(\overline{u}_{jh}\chi_m'\right) + 
    \frac{\partial}{\partial x_{jh}}\left(u_{jh}'\chi'\right)_m
\end{equation}

These operations leave a residual when comparing the region-averaged budgets to the first term on the right-hand side of eq.~\ref{e:reynoldstrans} due to $u^e_{jh}$. This is the term we dub ``net region expansion'' in the main. Since this residual also includes errors from numerical integration of simulation output and Reynolds averaging, it is the least well-constrained term in our budgets, but we find it plausible to attribute its main vertical structure to net expansion of the moist regions at the expense of dry regions.

\label{app:b}
\section{Derivation of evolution equation for $\partial/\partial z\left(\Gamma_{q_t} / \Gamma_{\theta_{lv}}\right)$}

To analyse the onset and evolution of mean-state convexity, i.e. $\partial^2 \overline{q_t}/\partial \overline{\theta_{lv}}^2>0$, we use the equivalence indicated in eq.~\ref{e:instabreq} and write an evolution equation for $\partial/\partial z\left(\Gamma_{q_t} / \Gamma_{\theta_{lv}}\right)$ by differentiating it to time and applying the quotient rule of calculus twice. This results in the following relation, where we have attempted to retain some brevity by writing (repeated) vertical derivatives as (repeated) subscripts $z$:

\begin{equation}
    \frac{\partial}{\partial t}\left(\frac{\overline{q_t}_z}{\overline{\theta_{lv}}_z}\right)_z =
    \frac{1}{\overline{\theta_{lv}}_z}\left(\frac{\partial \overline{q_t}}{\partial t}\right)_{zz}    -\frac{\overline{\theta_{lv}}_{zz}}{\overline{\theta_{lv}}_z^2}\left(\frac{\partial \overline{q_t}}{\partial t}\right)_z
    -\frac{\overline{q_t}_z}{\overline{\theta_{lv}}^2_z}\left(\frac{\partial\overline{\theta_{lv}}}{\partial t}\right)_{zz}
    +\left(2\frac{\overline{q_t}_z\overline{\theta_{lv}}_{zz}}{\overline{\theta_{lv}}_z^3} - \frac{\overline{q_t}_{zz}}{\overline{\theta_{lv}}_z^2} \right) \left(\frac{\partial \overline{\theta_{lv}}}{\partial t} \right)_z
    \label{e:gradtendnosub}
\end{equation}

To determine which processes influence the left-hand side of this equation, we may substitute for the tendencies that appear on its right-hand side terms from the budget for slab-averaged scalars eq.~\ref{e:slabav}. Applying the vertical derivatives term by term and inserting in eq.~\ref{e:gradtendnosub} results in eq.~\ref{e:gradtend}:

\begin{multline}
    \frac{\partial}{\partial t}\left(\frac{\overline{q_t}_z}{\overline{\theta_{lv}}_z}\right)_{z} = 
    \frac{1}{\overline{\theta_{lv}}_z}\left(
    - \left(\overline{w_{ls}} \overline{q_t}_z\right)_{zz}
    - \frac{1}{\rho_0}\left(\rho_0\overline{w'q_t'}\right)_{zzz}
    + \overline{S_{q_t}}_{zz}
    \right) \\
    -\frac{\overline{\theta_{lv}}_{zz}}{\overline{\theta_{lv}}_z^2}\left(
    - \left(\overline{w_{ls}} \overline{q_t}_z\right)_{z}
    - \frac{1}{\rho_0}\left(\rho_0\overline{w'q_t'}\right)_{zz}
    + \overline{S_{q_t}}_{z}
    \right) \\
    - \frac{\overline{q_t}_z}{\overline{\theta_{lv}}_z^2}\left(
    - \left(\overline{w_{ls}} \overline{\theta_{lv}}_z\right)_{zz}
    - \frac{1}{\rho_0}\left(\rho_0\overline{w'\theta_{lv}'}\right)_{zzz}
    + \overline{S_{\theta_{lv}}}_{zz}
    \right)  \\
    + \left(2\frac{\overline{q_t}_z\overline{\theta_{lv}}_{zz}}{\overline{\theta_{lv}}_z^3} - \frac{\overline{q_t}_{zz}}{\overline{\theta_{lv}}_z^2} \right)\left(
    - \left(\overline{w_{ls}} \overline{\theta_{lv}}_z\right)_{z}
    - \frac{1}{\rho_0}\left(\rho_0\overline{w'\theta_{lv}'}\right)_{zz}
    + \overline{S_{\theta_{lv}}}_{z}
    \right)
    \label{e:gradtend}
\end{multline}

Eq.~\ref{e:gradtend} highlights a few interesting requirements for the development of convexity in the mean state. First, it shows that in the limit of linear mean profiles ($\overline{\theta_{lv}}_{zz} = \overline{q_t}_{zz}=0$), the second and fourth term in eq.~\ref{e:gradtend} are zero, constraining the responsibility for the onset of convexity development in our simulations to processes that have curvature in their mean profiles (those in terms 1 and 3). However, the subsidence and large-scale forcing profiles of BOMEX are initially at most \textit{linear} functions of height. Therefore, the only non-zero terms that remain in eq.~\ref{e:gradtend} are third derivatives of the slab-mean fluxes. In fact, for $\Gamma_{\theta_{lv}}>0$ and $\Gamma_{q_t}<0$, these third derivatives must be negative to initiate the development of the required convexity, i.e. $\frac{\partial}{\partial t}\left(\frac{\overline{q_t}_z}{\overline{\theta_{lv}}_z}\right)_{z}>0$. The result is the condition eq.~\ref{e:divcurv} in the main text.

If $\overline{\theta_{lv}}$ and $\overline{q_t}$ have curvature in their profiles, the second and fourth terms are no longer necessarily zero, such that linear variations in subsidence and large-scale forcing, as well as curvature in the flux profiles, may have an effect.

\bibliographystyle{apalike}  
\bibliography{references}

\end{document}